\documentclass[aps,twocolumn,epsf,floats,prx,nofootinbib]{revtex4-1}
\usepackage{graphics,graphicx,epsfig}
\usepackage{xcolor}
\usepackage{amsmath}
\usepackage{amssymb}
\usepackage{bbold}
\usepackage{graphicx}
\graphicspath{{Figures/}}
\usepackage{bm}
\usepackage{verbatim}

\DeclareMathSymbol{\shortminus}{\mathbin}{AMSa}{"39}
\DeclareMathOperator*{\argmax}{arg\,max}

\newcommand{\beq}{\begin{equation}}
\newcommand{\eeq}{\end{equation}}
\newcommand{\<}{\langle}
\renewcommand{\>}{\rangle}

\newcommand{\ch}{}

\begin{document}

\title{Non-equilibrium dynamics of adaptation in sensory systems}

\author{Daniele Conti}
\affiliation{Laboratoire de physique de l'\'Ecole normale sup\'erieure, CNRS, PSL Universit\'e, Sorbonne Universit\'e, Universit\'e de Paris, 75005 Paris, France}
\author{Thierry Mora}
\affiliation{Laboratoire de physique de l'\'Ecole normale sup\'erieure, CNRS, PSL Universit\'e, Sorbonne Universit\'e, Universit\'e de Paris, 75005 Paris, France}

\begin{abstract}
Adaptation is used by biological sensory systems to respond to a wide range of environmental signals, by adapting their response properties to the statistics of the stimulus in order to maximize information transmission. We derive rules of optimal adaptation to changes in the mean and variance of a continuous stimulus in terms of Bayesian filters, and map them onto stochastic equations that couple the state of the environment to an internal variable controling the response function. We calculate numerical and exact results for the speed and accuracy of adaptation, and its impact on information transmission.
We find that, in the regime of efficient adaptation, the speed of adaptation scales sublinearly with the rate of change of the environment. Finally, we exploit the mathematical equivalence between adaptation and stochastic thermodynamics to quantitatively relate adaptation to the irreversibility of the adaptation time course, defined by the rate of entropy production. Our results suggest a means to empirically quantify adaptation in a model-free and non-parametric way.

\end{abstract}

\maketitle

\section{Introduction}
To make informed decisions, biological organisms sense and internally represent their environment using sensory systems, often with accuracy approaching physical limits \cite{Berg1977,Rieke:1998p6699}.
The {\em response function} of a sensory system maps input stimuli onto output signals, whose nature depends on the encoding system. For example, cells respond to environmental stimuli through biochemical signaling \cite{Berg,Macnab1972,Cheong2011}. The fate of cells in development is controled by patterns of gene expression \cite{Tkacik2011a}. Neural systems encode information using spikes, voltage differences, and ionic currents \cite{frazor2006local, wark2007sensory, webster2015visual}.

A common challenge posed to these sensory devices is the breadth of variation of external stimuli, which is often much larger than the response range of the sensory system. The activity of photoreceptors in the retina saturate over 2 orders of magnitude, yet they must cope with light intensities spanning 10 decades \cite{rieke2009challenges}. {\em E. coli} can navigate gradients of chemo-attractant concentrations over 5 orders of magnitude using the binary response of its rotary motors \cite{Lazova2011a}.
This is possible because over short timescales, stimuli are typically restricted to a much narrower distribution which depends on the immediate surrounding environment ---\,e.g. ambient light level in vision, or local concentration in chemical sensing. To produce efficient responses, sensory systems must adapt their response properties, as changes in the environment modify the statistical properties of the stimulus \cite{webster2015visual,Barkai1997}.

Theories has been developed to understand general principles of sensory adaptation and predict response properties, in particular in the context of sensory neuroscience \cite{weber2019coding}. One central idea is Barlow's efficient coding hypothesis \cite{attneave1954some, barlow1961possible}, which posits that neural systems maximize information transimission under the constraints of metabolic costs, dynamic range, and internal noise \cite{atick1992could, wainwright1999visual, bialek2006efficient,Tkacik2016}. Efficient coding predicts that the dynamic range of the response function should be matched to the distribution of inputs, so as to make the output distribution as balanced as possible, thus maximizing information \cite{laughlin1981simple,Callan2008}. As the distribution of input signals changes with time, the response function should then adapt accordingly. This argument rationalizes adaptation and makes specific predictions about its properties, with successful applications in vision \cite{brenner2000adaptive,Fairhall2001,wark2009timescales,weber2019coding}.
However, progress towards a general theory is hindered by the multiplicity of systems which differ in their constraints, costs, and relevant features to be encoded \cite{park2017bayesian, mlynarski2018adaptive}.

A possible overarching principle of sensory adaptation lies in its analogy to nonequilibrium statistical mechanics. When a thermal system is driven out of equilibrium by an external forcing, its energy landscape evolves and the system ``adapts'' by following the new equilibrium with a delay, while losing heat to the reservoir. The resulting dynamics are irreversible, and the dissipated work can be estimated using the measure of entropy production, which quantifies irreversibility \cite{Crooks1998}.
Similarly, sensory adaptation creates irreversible dynamics that carry an intrinsic energetic cost, as was studied in the case the {\em E. coli} chemotactic signaling network \cite{Lan2012}. {\ch The main result of this paper is to} formalize the link between adaptation and irreversibility by studying in detail analytically solvable systems of sensory adaptation, where the mean or the variance of the stimulus change over time.

{\ch We first derive the dynamics of optimal adaptation from principles of Bayesian inference (Sec.~II). While previous approaches have used Bayesian estimates to explain adaptation \cite{deweese1998asymmetric, wark2009timescales, mlynarski2018adaptive,Mynarski2019}, its impact on the response function were either not made explicit, or was obtained by minimizing a loss function reflecting particular coding constraints. By contrast, we derive the optimal response function from the maximization of Shannon's mutual information \cite{Tkacik2016}.
  A previous limitation of Bayesian update rules is that they usually assume discrete time, while both stimulus and response are continuous.
  We show that in the case of a varying mean, the adaptation dynamics have a well-defined continuous-time limit, which can be mapped onto a system of coupled Langevin equations with non-linear forces and position-dependent diffusivities, allowing for an explicit analogy with non-equilibrium statistical mechanics (Sec.~III).
We derive this mapping for the classical case of a stimulus mean switching between two values, but also for the case of a stimulus whose mean follows a random walk, which admits an analytical solution.
 In that continuous limit, we obtain explicit expressions for two quantities that characterize adaptation --- accuracy and speed --- which are in a trade-off relationship (Sec.~IV). To relax the assumption of optimality, we also explore performance in the case where the dynamical rules of the stimulus statistics are not precisely known (Sec.~V). The case of variance switching, which was treated in \cite{deweese1998asymmetric}, is revisited with predictions on how information transmission should drop and recover following switching events (Sec.~VI). Finally, we generalize and formalize the analogy between sensory adaptation and non-equilibrium statistical mechanism by proposing entropy production --- a non-parametric measure of temporal irreversibility --- as a signature of adaptation (Sec.~VI). We calculate it explicitly in the case of adaptation to a changing mean, and discuss its general relevance to experimental recordings of sensory neurons.
}

\begin{figure*}[htbp]
\centering
\includegraphics[width=1.0\textwidth]{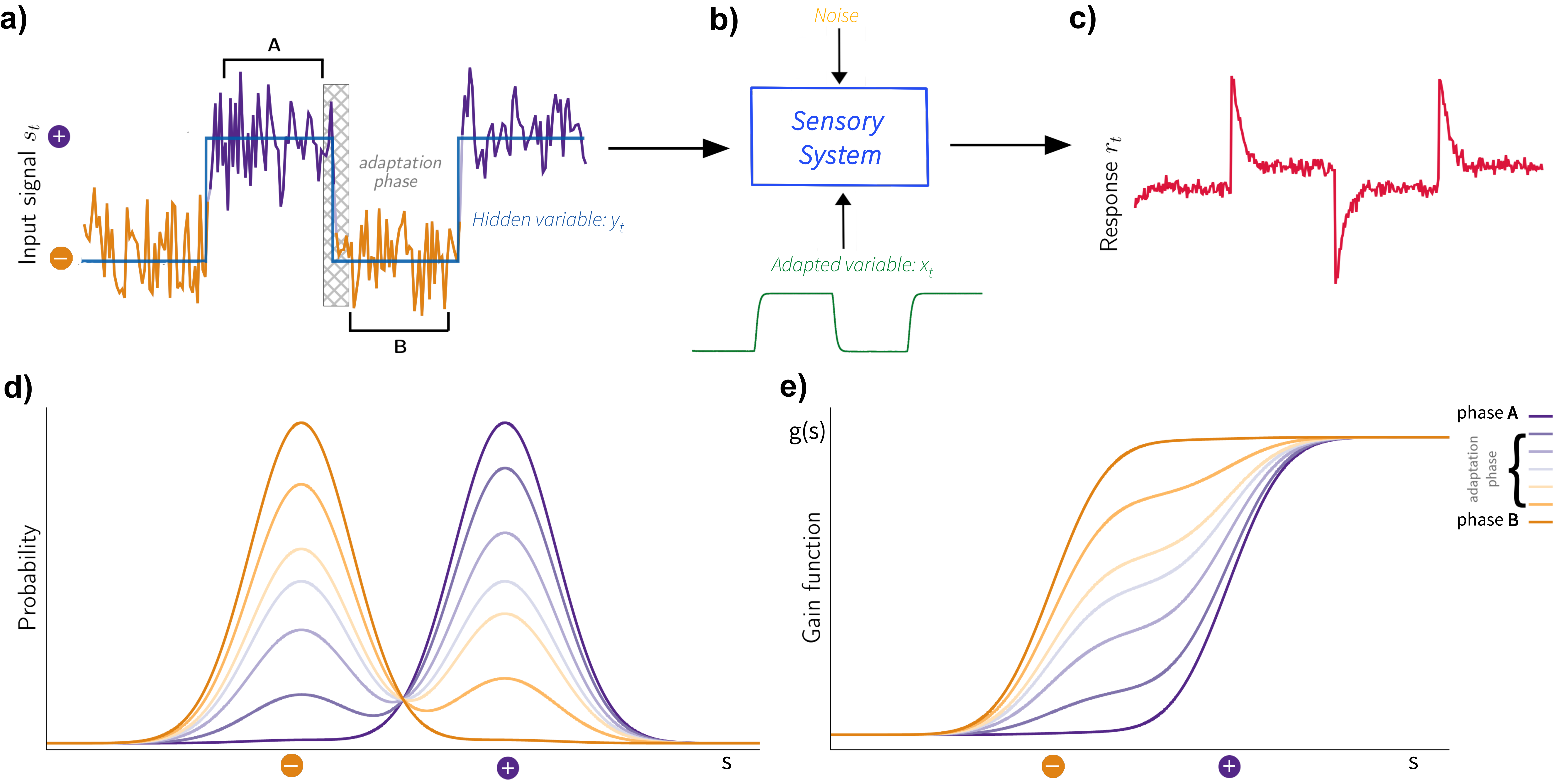}
\caption{
  \textbf{Sketch of the adaptation process}. \textbf{(a)} Example of a stochastic stimulus $s_t$ whose mean varies according to two environmental states {\ch $y_t=+$ and $-$}. After sufficient time in each state, the gain function of the sensory system is adapted to its statistics (phase A for $+$ state, and phase B for $-$ state). In addition, after each switch there is a transient phase during which the system adapts.
  {\ch {\bf (b)} The sensory system combines the raw stimulus $s_t$ with an internal estimate $x_t$ of the environmental variable $y_t$, computed from the past of $s_{t'\leq t}$.}
    \textbf{(c)} The sensory response {\ch $r_t$, which is optimized to maximize information transmission based on the internal variable $x_t$,} reacts to sudden changes of the mean stimulus, but then relaxes back to a basal value as the gain function adapts to the new state.
\textbf{(d)} Changes in the gain function can be rationalized based on the internal representation of the expected stimulus statistics (distribution of inputs), {\ch encoded by $x_t$}. When deep in the adapted phases A and B, {\ch the assumed distribution of inputs} matches the true statistics in each state (blue and orange curves {\ch for $x_t\approx +1$ and $x_t\approx -1$ respectively}). In the transient adaptation phase, new stimuli from the new state challenge the previous belief, which progresses to the new adapted phase (intermediate colors). \textbf{(e)} Information theory dictates that the optimal gain function should be most sensitive where the stimulus is expected (Eq.~\ref{eq:respfun}), causing the system to shift its dynamic range around the new mean. {\ch This explains the form of the response in (c).}
}
\label{fig:process_sketch}
\end{figure*}

\section{Adaptation as inference}
The first requirement to do efficient coding is to build a good estimate of the statistics of the environment, e.g. the mean and variance of the stimulus. To do so, the sensory system only has access to stimuli experienced in the past. It must integrate stimulus information far enough into the past to collect enough statistics, but not too far, since stimulus statistics themselves may change over time. This task can be naturally framed as an ongoing inference problem \cite{lochmann2011neural, lochmann2012perceptual, wark2009timescales, mlynarski2018adaptive}. Its optimal solution is given by a Bayesian formulation, which estimates the distribution of possible stimulus parameters given the stimulus history \cite{stocker2006sensory, wark2009timescales, park2017bayesian}.
{\ch Our approach postulates that the sensory system has access to the raw stimulus, rather than just the sensory response, to implement its adaptation mechanism. In this view, the sensory system ``compresses'' information by transforming the input into a noisy response, but can adapt this compression scheme based on the stimulus statistics, which it has access to through (unaltered) past stimuli.   This is particularly relevant for the early processing of visual information, which mostly proceeds in a feed-forward way (and, in fact, exclusively so in the retina). For instance, adaptation to the mean light level occurs directly at the level of photoreceptors, with no feedback from the retinal ganglion cells (the output of the retinal), and similarly for contrast adaptation at the level of signal transduction between photoreceptors and bipolar cells \cite{Rieke2001}.
  This view stands in contrast with adaptation mechanims through feedback, whereby the response function is adapted based on the knowledge of past sensory responses only \cite{mlynarski2018adaptive}.
}

The optimal Bayesian estimator \cite{deweese1998asymmetric,wark2009timescales,pfister2010synapses}, also called Bayesian filter, is defined as follows. 
At each time step $n$, the model estimates the state of the environment given the stimulus history, $P(y_n|s_{j\le n})$, where $y_n$ represents the state of the environment at time $n$, and $s_{j\le n}$ is the vector of past stimuli up until time $n$. In general $y_n$ parametrizes the distribution of stimuli at each time, $P(s_n|y_n)$. In this paper, this distribution will be assumed to be Gaussian, and $y_n$ will then simply denote its mean or variance.
Bayes's rule states that
\begin{equation}
    P(y_n| s_{j \le n}) = \frac{1}{\Omega} P(s_n|y_n) P(y_{n}|s_{j <n}),
    \label{fullbayes}
\end{equation}
where $\Omega = P(s_{j \le n})$ normalizes the distribution.

Assuming that the dynamics of the environment is Markovian, one can write a recursive form:
\begin{equation}
    P(y_n| s_{j \le n}) = \frac{1}{\Omega}P(s_n|y_n) \sum_{y_{n\shortminus1}} P(y_n|y_{n\shortminus1})P(y_{n\shortminus1}|s_{j <n})
    \label{fullbayes}
\end{equation}
This formula combines the new observation $s_n$ with the estimate of $y_{n\shortminus 1}$ at the previous time step \cite{deweese1998asymmetric}, taking into account the way $y_n$ may have evolved. Prior knowledge on $P(y_n|y_{n\shortminus 1})$ is essential as it sets the time scale over which past samples are discarded. Ignoring it, as was done in \cite{wark2009timescales}, leads to assuming an infinite memory timescale and everlasting dependence on initial conditions.

Armed with an estimate of the environment statistics, the sensory system may use this knowledge to adapt its response $r_n$ to the stimulus. 
Following previous proposals \cite{Barlow1961,Bialek_spikes,brenner2000adaptive,Callan2008}, one may assume that the stochastic encoding $P(r_n|s_n)$ is chosen to maximize information transmission:
\beq\label{eq:MI}
I(s_n,r_n)=\int ds_n\, dr_n\, P(s_n)P(r_n|s_n)\ln\frac{P(r_n|s_n)}{P(r_n)},
\eeq
where the form of the response function $P(r_n|s_n)$ is set by biophysical constraints.
{\ch Alternative choices of objectives to optimize include decoding accuracy, metabolic costs \cite{Levy1996}, or the ability to infer the environmental variable $y_n$ itself \cite{mlynarski2018adaptive}.}
The simplest assumption, which is equivalent to Laughlin's original argument \cite{laughlin1981simple}, is to assume a constant Gaussian output noise:
\beq
r_n=g(s_n) + \epsilon_n,
\eeq
where $\epsilon_n$ is a uncorrelated Gaussian noise of zero mean and variance $\sigma_\epsilon^2$, and $g(s)$ is a function constrained between $0$ and $r_{\rm max}$, so that $P(r|s)\propto \exp[-(r-g(s))^2/2\sigma_{\epsilon}^2]$. In the small noise limit $\sigma_\epsilon\ll g_{\rm max}$, the response function $g(s)$ maximizing the mutual information is the one that maximizes the entropy of the response \cite{Callan2008}, which is realized by the uniform distribution $P(r)=1/r_{\rm max}$ for $0\leq r\leq r_{\rm max}$, which in turns gives $P(r)dr\approx \hat P(s)ds$, hence $dr/ds=g'(s)=r_{\rm max}\hat P(s)$, or equivalently
\beq\label{eq:respfun}
g(s)=r_{\rm max}\int_{-\infty}^s ds' \hat P(s'),
\eeq
so that the response function's sensitivity follows the assumed statistics $\hat P$ of the stimulus at time $n$, derived from the belief about the state of the environment $y$:
\beq\label{eq:belief}
\hat P(s)=\int dy\, P(s|y)P(y|s_{j\leq n}).
\eeq
The mutual information being actually transmitted can be rewritten as:
\beq
I(s,r)=S[r]-S[r|s]
\eeq
with $S[r]=-\int_{r} dr P(r)\ln P(r)$, and $S[r|s]=-\int ds P(s|y)\int P(r|s)\ln P(r|s)$, where $P(r)=\int ds\,P(r|s)P(s|y)$. For Gaussian output noise, the second term is just the entropy of the noise, $S[r|s]=(1/2)\ln(2\pi e \sigma_\epsilon^2)$. In the same limit of small noise that gaves us the optimal encoding, the output is almost a deterministic function of the input, so that $S[r]$ is approximately obtained using $P(r)\approx P(s|y)ds/dr=r_{\rm max}^{-1}P(s|y)/\hat P(s)$, proportionnal to the ratio of the true to the assumed stimulus distributions. This resulting information transmitted is:
\beq\label{eq:Is}
I(s,r) = \frac{1}{2}\ln\frac{2\pi er_{\rm max}^2}{\sigma_\epsilon^2} - \int ds\, P(s|y)\ln\frac{P(s|y)}{\hat P(s)},
\eeq
where in the second term we recognize a Kullback-Leibler divergence or cross-entropy, with is always non-negative.
The mutual information is maximized when the inferred statistics of the environment exactly matches the true one, $\hat P(s)=P(s|y)$. {\ch In that case, the Kullback-Leibler divergence is zero, and only the first term survives --- the channel capacity --- which has an intuitive interpretation: it is the logarithm of the number of distinct responses that can be resolved from each other, given by the ratio of the output dynamic range, $r_{\rm max}$, to the resolution of the response, $\sigma_\epsilon$. This number, which can also be interpreted as a signal-to-noise ratio, encodes the resource constraints of the sensory system, and is the main determinant of information transmission. However, an interesting feature of \eqref{eq:Is} is that the information loss due to non-optimal adaptation encoded in the Kullback-Leibler divergence does {\em not} depend on that resource parameter, making the analysis robust to that choice.
}

The full adaptation scheme is summarized in Fig.~\ref{fig:process_sketch} with the example of a Gaussian stimulus whose mean switches between two values. Shortly after the mean changes (Fig.~\ref{fig:process_sketch}a), the system progressively updates its expected distribution of stimuli as it accumulates more samples (Fig.~\ref{fig:process_sketch}d), and adapts its response function (Fig.~\ref{fig:process_sketch}b) according to \eqref{eq:respfun}, as illustrated in Fig.~\ref{fig:process_sketch}e. The resulting response dynamics (Fig.~\ref{fig:process_sketch}c) shows epochs of strong upward and downward changes right after each switching transitions, followed by relaxation to intermediate values as the system adapts to its new response function, closely mimicking experimental observations.

In the following we will study and solve particular cases of adaptation to a Gaussian stimulus with varying mean or variance.

\section{Adaptation to a varying mean}

Let us start with the most intuitive and simple case: adaptation to changes in the mean of the stimulus.
We will take two very simple processes as case studies: a telegraph process,
in which the mean randomly switches between two values, and an
Ornstein-Uhlenbeck process, in which the mean varies continuously
around a fixed value. Discontinuous switches such as produced by the
telegraph process are commonly studied in
experiments on sensory adaptation, because of the strong adaptive transient they trigger. The
Ornstein-Uhlenbeck process is interesting because of the analytical insight it affords.

\subsection{Two models of fluctuating mean}
\begin{figure*}[htbp]
\centering
\includegraphics[width=\textwidth]{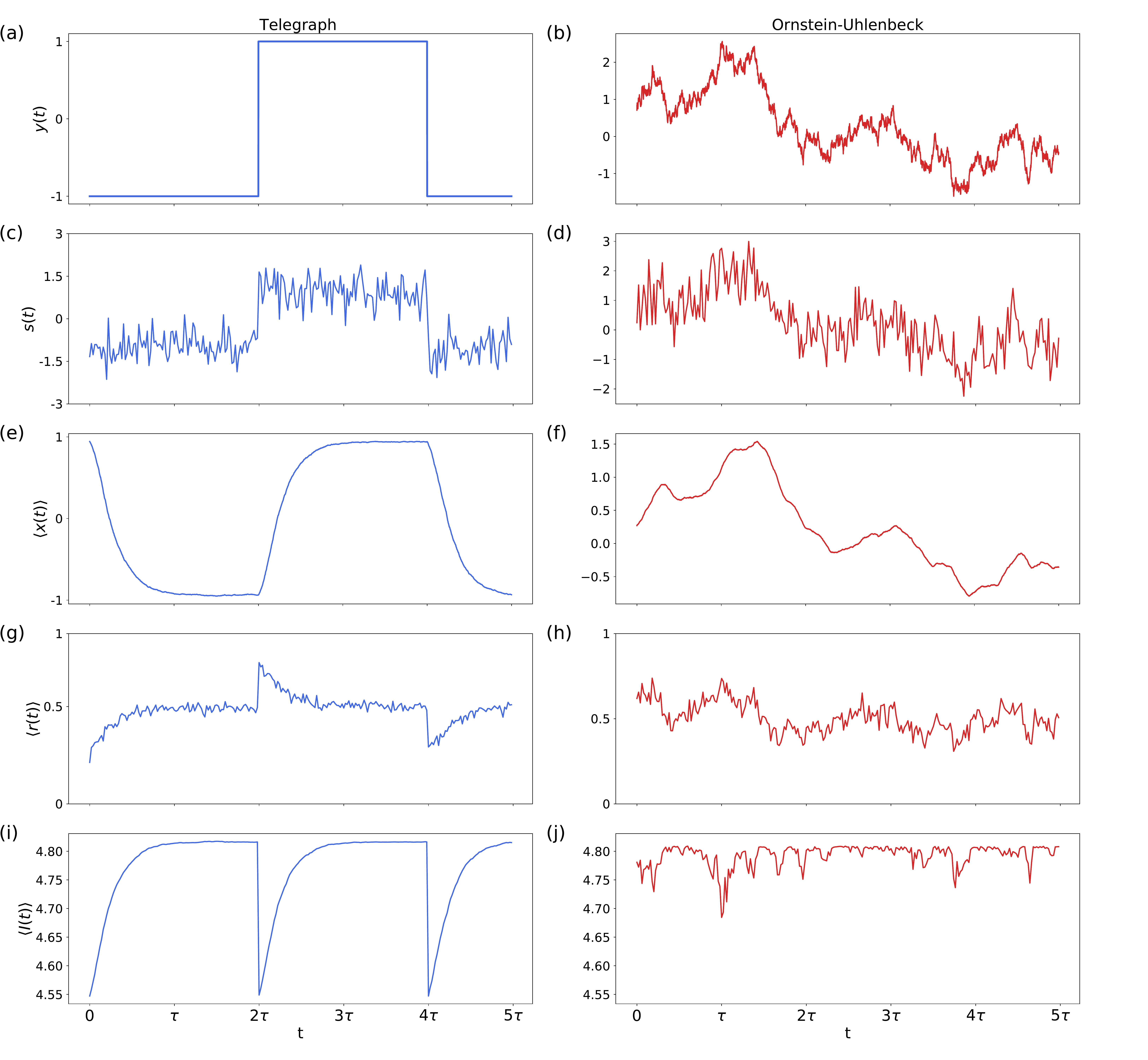}
\caption{
\textbf{Adaptation dynamics for two mean-varying processes: telegraph process (left), and Ornstein-Uhlenbeck process (right).} 
\textbf{(a),(b)} The state variable $y(t)$---the mean of the stimulus---varies over time with unit variance and correlation timescale $\tau$.
\textbf{(c),(d)} Perceived stimulus $s(t)=y(t)+\sqrt{\theta}\xi(t)$.
\textbf{(e),(f)} Mean value of the state variable, $x=\int dy\,yP(y|\{s(t'\leq t)\}$) inferred by the system from past observations.
\textbf{(g),(h)} Sensory response obtained with the optimal gain function $r=g(s)$, derived by optimizing information transmission.
\textbf{(i),(j)} Information \eqref{eq:MI} actually transmitted per time bin ($\delta t=10^{-3}$, $\sigma_{\epsilon} =0.01$).
Brackets denote averages over 500 repetitions of the state variable evolution.
}
\label{fig:mean_example}
\end{figure*}

\subsubsection{Telegraph process}

We consider a system where the environmental parameter $y_n$ switches
between two values $y_-,y_+$ with probability $k$ at each time step
The sensory system has access only to the $s_n$, which we assume to be Gaussian distributed with mean $y_n$ and fixed variance $\sigma^2$,
\begin{equation}
  s_n = y_n + \sigma \eta_n,
\label{eq:stimulus}
\end{equation}
with $\langle \eta_n \rangle = 0$ and $\langle \eta_n \eta_{n'} \rangle = \delta_{nn'}$, so that:
\begin{equation}
P(s_n|y_n) = \frac{1}{\sqrt{2 \pi \sigma^2}}e^{-\frac{(s_n - y_n)^2}{2 \sigma^2}}
\label{like_mean}
\end{equation}

We replace this expression into \eqref{fullbayes}, and use the fact that
the distribution of $y_n$ is entirely determined by its mean $x_n
\equiv \<y_n\>_{|s_{j \le n}}$ because of its binary nature. Assuming
$y_{\pm} = \pm 1$ without any loss of generality, by averaging
\eqref{fullbayes} we then obtain an update equation for $x_n$:
\begin{equation}
x_n = \frac{\sinh(\frac{s_n}{\sigma^2}) + (1-2k)x_{n\shortminus1}\cosh(\frac{s_n}{\sigma^2})}{\cosh(\frac{s_n}{\sigma^2}) + (1-2k)x_{n\shortminus1}\sinh(\frac{s_n}{\sigma^2})}.
\label{switch_fullx}
\end{equation}

For this estimator to be optimal, the variance $\sigma^2$ in \eqref{switch_fullx} should be equal to the true variance of the stimulus distribution.
This can be achieved, for instance, by a variance adaptation mechanism
as we will see in Sec.~\ref{sec:variance}.
In Sec.~\ref{sec:maladapt} we will also deal with the case where the system does not use optimal parameters.

\subsubsection{Auto-regressive (Ornstein-Uhlenbeck) process}

Another simple example of stochastic dynamics for the mean is given by
an auto-regressive process, which we also call Ornstein-Uhlenbeck
process by abuse of language, as it reduces to it in the continuous
time limit (see later):
\begin{equation}\label{eq:auto}
    y_{n+1} = ay_n + \sigma' \eta'_n
\end{equation}
where $\eta'$ is a Gaussian white noise, $\langle \eta'_n \rangle = 0$
and $\langle \eta'_n \eta'_{n'} \rangle = \delta_{nn'}$, and where $a<1$.
Again we assume that signals are distributed according to a
Gaussian of mean $y_n$ and variance $\sigma^2$, $s_n = y_n + \sigma
\eta_n$, with $\langle \eta_n \rangle = 0$ and $\langle \eta_n
\eta_{n'} \rangle = \delta_{nn'}$, $\<\eta'_n\eta_{n'}\>=0$.

Since the elements of the recursion are Gaussian, we assume a Gaussian Ansatz for the posterior:
\begin{equation}
    P(y_n|s_{j\le n}) = \frac{1}{\sqrt{2 \pi u_n^2}} \, e^{- \frac{(y_n - x_n)^2}{2 u_n^2}} \,.
    \label{posterior_ou}
\end{equation}
Plugging it into \eqref{fullbayes} yields recursive equations for the
mean and variance of the posterior of $y_n$, which are equivalent to a
Kalman filter \cite{Kalman1960-kf}:
\begin{align}
  x_n &= \frac{ a\sigma^2 x_{n\shortminus1} + \left[\sigma'^2 + a^2u_{n\shortminus1}^2\right] s_n}{ \sigma^2 + \sigma'^2 + a^2u_{n\shortminus1}^2}\label{ou_discrete} \\
    u_n^2 &= \frac{\sigma^2\left[\sigma'^2 + a^2u_{n\shortminus1}^2\right]}{\sigma^2 + \sigma'^2 + a^2u_{n\shortminus1}^2}.
\label{ou_discrete_var}
\end{align}
The dynamics of the variance \eqref{ou_discrete_var} does not depend
on the mean or the stimulus, so that it converges to a fixed
point $u^2$ at steady state.

\subsection{Continuous time limit}

Our adaptation dynamics were derived at discrete times, using a Bayesian estimator and exploiting the Markovian property of the system.
We now take the limit of continuous time to obtain stochastic differential equations. Denoting by $\delta t$ the time step between two observations, the continuous time is defined as $t=n\delta t$, with $\delta t\to 0$. As we take this limit, we must scale the various parameters of the dynamics to ensure that all the relevant quantities remain finite.

For both types of environmental dynamics, the stimulus variance may scale with $\delta t$ as:
\beq
\sigma^2 = \frac{\theta}{\delta t}.
\label{noise_scaling}
\eeq
This scaling allows us to write the stochastic term in \eqref{eq:stimulus} as the discretization of a Gaussian white noise $\xi(t)$:
$\sigma \eta_n = (\sqrt{\theta}/{\delta t}) \int_{t_n}^{t_n + \delta t} dt \, \xi(t)$,
with $\<\xi(t)\>=0$ and $\<\xi(t)\xi(t')\>=\delta(t-t')$. In the
continuous time limit this gives:
\beq\label{eq:signal}
s(t)=y(t)+\sqrt{\theta}\xi(t).
\eeq

In the telegraph process, the switching probability between two infinitesimal time steps should scale with $\delta t$. We define
$k = {\delta t}/({2 \tau})$.
The switching rate is $(2\tau)^{-1}$, so that $\tau$ is the correlation timescale of the process.

Similarly, a continuous Ornstein-Uhlenbeck (OU) process
may be obtained from the auto-regressive process \eqref{eq:auto} in the $\delta t\to 0$ limit with the scalings
  $a=1-\delta t/\tau$,
$\sigma'^2=2D\delta t$:
\beq\label{eq:ou}
\frac{dy}{dt}=-\frac{y}{\tau}+\sqrt{2D}\xi'(t),
\eeq
with $\xi'(t)$ a Gaussian white noise. $\tau$ is a relaxation time, and $D$ may be interpreted as a diffusion coefficient in stimulus space. We set it to $D=1/\tau$, so that the variance of $y$ is 1.
With these choices, the first two moments of $y(t)$ are the same for both the telegraph and OU processes,
\beq\label{eq:moments}
\<y(t)\>=0,\qquad\<y(t)y(t')\>=e^{-|t-t'|/\tau}.
\eeq

With these scalings, the adaptation dynamics are described in the $\delta t\to 0$ limit by stochastic differential equations (with It\^o convention). For the telegraph process, \eqref{switch_fullx} becomes:
\begin{equation}
\frac{dx}{dt} = -\frac{x}{\tau} + \frac{1-x^2}{\theta}s= -\frac{x}{\tau} + \frac{1-x^2}{\theta} y + \frac{1-x^2}{\sqrt{\theta}} \xi.
\label{continuous_x}
\end{equation}
For the OU process, \eqref{ou_discrete} becomes:
\begin{equation}
  \begin{split}
    \frac{d x}{dt} &= -\frac{\sqrt{1 + 2\tau/\theta}}{\tau}x +\frac{u^2}{\theta} s\\
    &= -\frac{\sqrt{1 + 2{\tau}/{\theta}}}{\tau}x +\frac{u^2}{\theta} y +\frac{u^2}{\sqrt{\theta}} \xi,
\end{split}
\label{ou_eq}
\end{equation}
where the posterior variance
\begin{equation}
u^2 = \mathrm{Var}(y(t)|\{s(t'\leq t)\})=\frac{2}{1+\sqrt{1+2\tau/\theta}}
\end{equation}
is obtained from the fixed-point condition
\eqref{ou_discrete_var}. The adaptation dynamics for the OU process
thus follow an exactly solvable Gaussian process described by the
coupled equations \eqref{eq:ou} and \eqref{ou_eq}.

Two timescales, $\tau$ and $\theta$, govern the adapation dynamics of both processes \eqref{continuous_x}-\eqref{ou_eq}: $\tau$ is the timescale over which the environment varies, while $\theta$ gives the typical timescale over which the state variable, $y$, and the fluctuations of the stimulus, $\sqrt{\theta}\xi$, have equal contributions: $\<[(1/T)\int_0^T \sqrt{\theta}\xi(t)dt]^2\>=\theta/T \sim \<y^2\>= 1$ for $T \sim\theta$.
The adaptation dynamics is expected to depend crucially on the ratio between these two timescales:
\beq
\alpha = \frac{\theta}{\tau},
\eeq
which can be viewed as an inverse signal-to-noise ratio, and is the
main control parameter of adaptation. A low $\alpha$ means low
stimulus fluctuations and thus precise adaptation, while high $\alpha$ means high fluctuations and thus poor adaptation.

The information transmitted under the optimal scheme described in the
previous section is given by \eqref{eq:Is} with
\beq
\hat P(s) = \frac{1}{\sqrt{2\pi
    (\sigma^2+u^2)}}\exp\left[-\frac{(s-x)^2}{2(\sigma^2+u^2)}\right]
\eeq
for the OU process,
and
\beq
\hat P(s) = \frac{1}{\sqrt{2\pi
    \sigma^2}}\left[\frac{1+x}{2}e^{-\frac{(s-1)^2}{2\sigma^2}}+\frac{1-x}{2}e^{-\frac{(s+1)^2}{2\sigma^2}}\right]
\eeq
for the telegraph process. In the OU case the expression for the information simplifies
to:
\beq\label{eq:IOU}
I(s,r)=\frac{1}{2}\ln\frac{2\pi er_{\rm max}^2}{\sigma_\epsilon^2}-\frac{1}{2}\ln\left[1+\frac{u^2}{\sigma^2}\right]-\frac{1}{2}\frac{(x-y)^2-u^2}{\sigma^2+u^2}.
\eeq

Fig.~\ref{fig:mean_example} shows numerical simulations of the optimal adaptation dynamics \eqref{continuous_x} and \eqref{ou_eq} for the two processes, as well as the expected sensory response and information rate under the assumption of optimal information transmission \eqref{eq:MI}-\eqref{eq:belief}. For the telegraph process, the response shows typical adaptive behaviour, with fast changes in the response following a switch, followed by a slower relaxation to the baseline. The information rate drop right after each switch, and climbs back up to its maximum as the system adapts its response function to the new statistics.

\section{Accuracy and speed of optimal adaptation}

To study the adaptation dynamics of \eqref{continuous_x} and
\eqref{ou_eq}, we focus on two fundamental properties: the
speed of adaptation, measured by the typical time it takes for the
system to adapt to the changing environment, and the accuracy of
adaptation, measured by the discrepancy between $x$ and $y$.

\begin{figure*}[t!]
\centering
\includegraphics[width=1.0\textwidth]{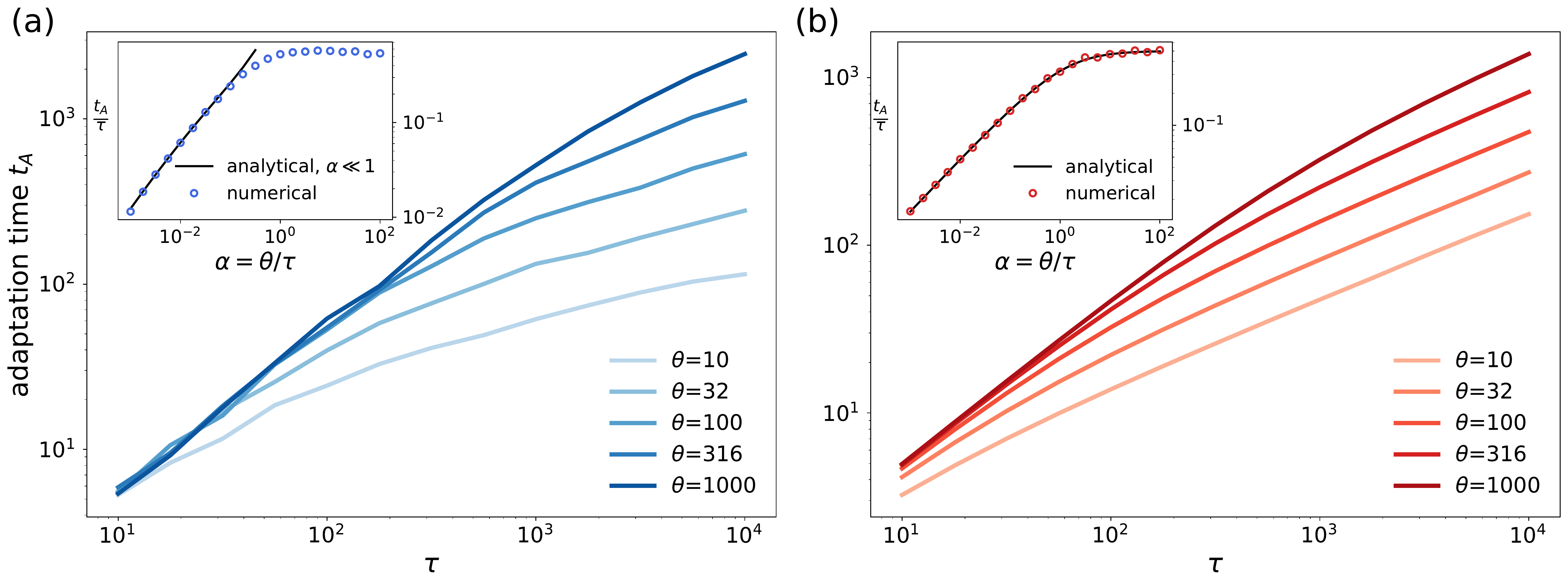}
\caption{
  \textbf{Time of adaptation to the mean stimulus, $t_A$.} The adaptation time is defined as the time at which the cross-correlation between the true and inferred state variable is maximum \eqref{eq:ta}. It is shown as a function of the environmental time scale $\tau$ for \textbf{(a)} the telegraph process and  \textbf{(b)} the Ornstein-Uhlenbeck process. The results are obtained numerically for the telegraph process, and analytically for the OU process \eqref{tau_ou_full}.
  Insets show the rescaled adaptation time $t_A/\tau$ as a function of the control parameter $\alpha$, for both simulations (circles), and analytical predictions (Eq.~\ref{tau_switch_full} for the telegraph process, valid for $\alpha\ll 1$, and Eq.~\ref{tau_ou_full} for the OU process).
}
\label{fig:mean_time}
\end{figure*}

\subsection{Adaptation time}
We formally define the adaptation time as the time $t=t_A$ at which
the cross-correlation function,
\beq
C(t)=\<y(t_0)x(t_0+t)\>,
\eeq
reaches its maximum,
\beq\label{eq:ta}
t_A={\argmax}_t  C(t).
\eeq
This defines the time
delay after which the adaptation variable, $x$, is maximally aligned
with the state variable, $y$, {\em i.e.} the
time it takes for $x$ to ``catch up'' with $y$.

We first compute this adaptation time numerically
(Fig.~\ref{fig:mean_time}) for both processes and for different values
of the control parameter $\alpha$, using the Euler–Maruyama method \cite{Kloeden}. In general, the adaptation
timescale $t_A$ grows with the timescale of switching, $\tau$. This
suggests that sensory systems should modify their dynamics of
adaptation as a function of what their expected rate of change
is. Such meta-adaptation of the relaxation timescale has indeed been observed in the context of
mean adaptation to the mean \cite{wark2009timescales}.

Two broad regimes can be distinguished. In
the low signal-to-noise regime, $\alpha\gtrsim 1$, $t_A$ scales
linearly with $\tau$: the optimal adaptation timescale
is proportional to the timescale of the environment. However, in the
well-adapted regime, $\alpha\ll 1$, this scaling breaks down, and
adaptation happens much faster than the rate of change of the environment would
suggest. In that regime, it's the reliability of the observed signals
that drives how fast the system adapts, so that small values of
$\theta$ lead to fast adaptation.

To gain analytical insight into these different scalings, we first
consider the OU process, for which the cross-correlation can be
calculated analytically:
\begin{equation}
    C( t)  = e^{-\frac{t}{\tau}}\left[1 -\frac{2}{1+\sqrt{1+2/\alpha}}e^{-\frac{(\sqrt{1 + 2/\alpha} -1)t}{\tau}}\right],
\end{equation}
yielding:
\begin{equation}
\frac{t_A}{\tau}= \frac{\alpha(1+\sqrt{1+2/\alpha})}{2} \ln{\left(\frac{2\sqrt{1+2/\alpha}}{1+\sqrt{1+2/\alpha}}\right)}.
\label{tau_ou_full}
\end{equation}
For large $\alpha$, this expression becomes at leading order $t_A\sim
\tau/2$, confirming the linear scaling observed in that regime. For small
$\alpha$, we get $t_A\sim \sqrt{\tau\theta/2}\ln(2)$, meaning that the
adaptation scale is the geometric mean of the environmental timescale
and the noise timescale. Such a scaling, which was previous derived in the
context of Bayesian filtering for concentration sensing
\cite{Mora2019}, results from a trade-off between the requirements
to integrate information about the stimulus for as long as possible
($t_A\gg \theta$), but not too long to avoid including out-of-date
information ($t_A\ll \tau$).

While the adaptation dynamics for the telegraph process doesn't have
an analytical solution, we may approximate $t_A$ by the typical time $x$ takes to cross 0
following a switch. This allows us to cast the problem as a first
passage calculation. We start by writing the Fokker-Planck equation
describing the density evolution of \eqref{continuous_x} as a function of time:
\beq
\frac{\partial\rho}{\partial t}=-\frac{\partial}{\partial
  x}\left[\left(-\frac{x}{\tau}+\frac{1-x^2}{\theta}y\right)\rho\right]+\frac{\partial^2}{\partial
  x^2}\left(\frac{(1-x^2)^2}{2{\theta}}\rho\right),
\eeq
whose steady-state solution at constant $y$ reads:
\beq\label{eq:eq}
\rho_{\rm
  eq}(x|y)=\frac{N}{(1-x^2)^2}{\left(\frac{1+x}{1-x}\right)}^y\exp\left(-\frac{\alpha}{1-x^2}\right),
\eeq
where $N$ is a normalization constant.

In the $\alpha\ll 1$ limit, $x$ is typically close to $y$ and its
difference with it is of order $\alpha$: $1-yx=\mathcal{O}(\alpha)$, as can be checked by solving $d\rho_{\rm eq}/dx=0$.
Suppose that at the time $t=0$ of a switch from $y=+1$ to $y=-1$,
$x=x_0$ is drawn from the steady-state distribution $\rho_{\rm
  eq}(x|y=+1)$ \eqref{eq:eq}. We define $\mathcal{T}(x_0)$ as the mean
first passage time of $x$ from $x_0$ to 0, with $x=+1$ acting as a reflecting
boundary. This time is given by
\cite{gardiner2009stochastic}:
 \begin{equation}
 \mathcal{T}(x_0) = 2\theta \int_{x_0}^0 \frac{dx}{\psi(x)}\int_{-1}^x
 dx' \frac{\psi(x')}{(1-x'^2)^2}
 \label{fpt}
 \end{equation}
where
\begin{equation}
\psi(x)={\left(\frac{1-x}{1+x}\right)}\exp\left(-\frac{\alpha}{1-x^2}\right).
\end{equation}
The adaptation time is then $t_A\approx \<\mathcal{T}(x_0)\>_{x_0\sim
  \rho_{\rm eq}(x|y=+1)}$.

Examining Eq.~\ref{continuous_x} in the small $\alpha$ regime, we see
that when $x$ is close to $+1$, the $(1-x^2)$ terms scales with
$\alpha$, meaning that $x$ has sluggish dynamics dominated by the time
scale $\tau$. When $x$ finally gets away from $+1$, the dynamics accelerates according to the fast time scale $\theta$, quickly reaching zero. This motivates us to use the change of
variable $x=1-\alpha /(2v)$, and expand in $\alpha$ to
obtain (see Appendix \ref{app:fpt}):
\begin{equation}\begin{split}
    \frac{t_A}{\tau}\simeq \frac{\alpha}{4}
    \left[\ln^2(\alpha)+(2\gamma_{\rm e}-2\ln(2)+1)\ln(\alpha)+C\right]+{o}(\alpha),
\label{tau_switch_full}
\end{split}\end{equation}
where $\gamma_{\rm e}$ is Euler's constant, $C\approx 7.59$ is a numerical constant. We checked
the validity of this expression against simulation results (Fig.~\ref{fig:mean_time}a).

The leading order in $\alpha$ scales as:
\begin{equation}
t_A\sim \frac{\theta}{4} \ln^2\left(\frac{\tau}{\theta}\right).
\end{equation}
This sublinear scaling with $\tau$ stands
in contrast with the theorical results of Wark {\em et al.} \cite{wark2009timescales},
which reported a linear one. This difference is explained
by the fact that in \cite{wark2009timescales}, the inference system assumed a
vanishing switching rate. This leads to the absence of a well defined
steady-state, requiring to evaluate the adaptation scale over
simulations of finite durations, with a dependence on initial
conditions.

This scaling is also much more faster with $\theta$ than
the square-root scaling obtained for the OU process. This is because
the OU process assumes that the mean changes all the time, instead of
by discontinuous jumps, making adaptation faster.

\begin{figure*}[t!]
\centering
\includegraphics[width=1.0\textwidth]{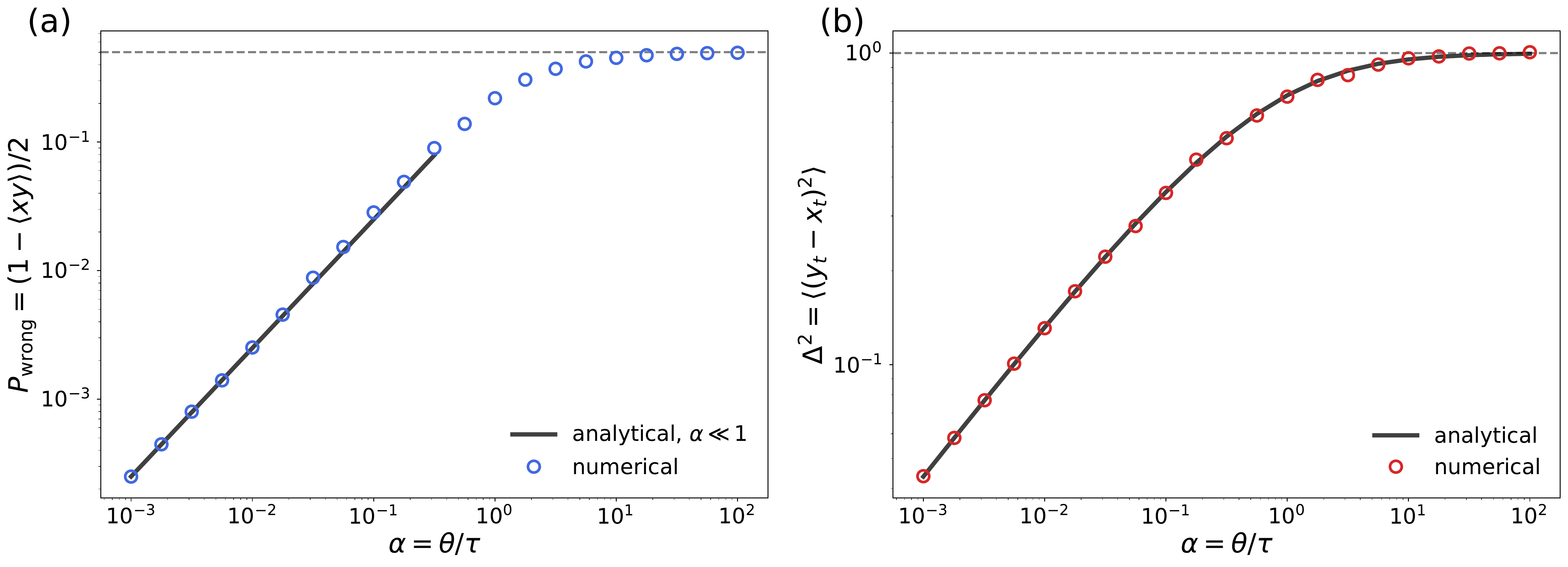}
\caption{
\textbf{Accuracy of adaptation to the mean stimulus as a function of rescaled stimulus variance $\alpha=\theta/\tau$.}
\textbf{(a)} Accuracy in the telegraph process, defined as the value of the probability of being in the wrong state $P_{\rm wrong}=(1-\<xy\>)/2$.
Circles are obtained by numerical simulation, and the line is the analytical approximation given by \eqref{accuracy_switch} for $\alpha\ll 1$.
\textbf{(b)} Accuracy in the Ornstein-Uhlenbeck process, defined as the posterior mean squared error $\Delta^2 = \langle(y-x)^2\rangle$.
Circles are from numerical simulations, and the line is the analytical prediction given by \eqref{accuracy_ou}. In both panels the dashed line shows chance prediction ($P_{\rm wrong}=1/2$ and $\Delta^2=1$)}
\label{fig:accuracy}
\end{figure*}

\subsection{Accuracy}

We introduce two definitions of accuracy adapted to each case.
For the telegraph process, which admits only two values for the state
variable $y=\pm 1$, we define the discrepancy as the probability for
the posterior to be wrong, $P_{\rm wrong}=(1-\<xy\>)/2$, long after
the adaptation transient.
For the OU process, whose the state variable is continuous, it is more
natural to define the discrepancy of adaptation as the standard error between $x$ and
$y$: $\Delta^2 = \langle (y - x)^2 \rangle$ {\ch (although note that $P_{\rm wrong}=\<|\hat
y-y|^2\>/4$, where $\hat y$ is drawn from the posterior $P(\hat
y|s_{t'\leq t})=\frac{1+x}{2}\delta(\hat y-1)+\frac{1-x}{2}\delta(\hat y+1)$, can also be
viewed as a mean-squared difference)}.

\subsubsection{Ornstein-Uhlenbeck}
Since the OU process is Gaussian, all moments can be calculated exactly.
We can rewrite $\Delta^2 = \langle (y - x)^2 \rangle=\langle y^2
  \rangle + \langle x^2 \rangle - 2\langle yx \rangle$.
We already know $\langle y^2 \rangle = 1$, and
calculate $\langle x^2 \rangle =\langle yx \rangle = 1 -
\alpha(\sqrt{1+2/\alpha}-1)$, from which we get
\begin{equation}\label{accuracy_ou}
\Delta^2 = \alpha\left(\sqrt{1+2/\alpha}-1\right)=u^2.
\end{equation}
(Note that this error is equal to the uncertainty computed by the Bayesian inference system, $u^2$, consistent with its optimality).
In Fig.~\ref{fig:accuracy}b, we plot this result and check its
validity by comparing it to numerical integration of \eqref{eq:ou}
and \eqref{ou_eq}.
In the limit of perfect adaptation, $\alpha \ll 1$, we get at
leading order
\begin{equation}\label{eq:OUsmallalpha}
\Delta^2 \sim (2\alpha)^{1/2}.
\end{equation}

The expression for the accuracy allows us to compute the average
information between stimulus and response \eqref{eq:IOU}.
\begin{equation}\label{eq:IOUavrg}
 I(s,r)\approx  \<I(s,r)\>=\frac{1}{2}\ln\frac{2\pi er_{\rm max}^2}{\sigma_\epsilon^2}-\frac{1}{2}\ln\left[1+\frac{u^2}{\sigma^2}\right].
\end{equation}
{\ch The comparison between numerics and this expression is
shown in Fig.~\ref{fig:information}b.
In the
$\alpha\ll 1$ regime, information approaches channel capacity, i.e. the maximal value allowed
by the output noise, $I(s,r)\approx I_{\rm max}=\frac{1}{2}\ln[2\pi e(r_{\rm max}/\sigma_\epsilon)^2]$.
}

\subsubsection{Telegraph process}
Because $P_{\rm wrong}$
is a steady-state property, it only depends on the control
parameter $\alpha$. We computed it by simulating a
telegraph process $y(t)$ with switching rate $1/(2\tau)$, and by
integrating \eqref{continuous_x} numerically
(Fig.~\ref{fig:accuracy}a).
For small $\alpha$, adaptation is very accurate, while for
$\alpha\gtrsim 1$ it is very poor, with $P_{\rm wrong}$ quickly
reaching chance level $\approx 1/2$.

In the well-adapted phase $\alpha\ll 1$ ($\tau\gg \theta$), adaptation
is fast compared to the switching of the state variable. After each
transition, following an adaptation transient, the system quickly reaches the
steady state of \eqref{continuous_x} with fixed $y$, given by \eqref{eq:eq}.
While the moments of $\rho_{\rm
  eq}(x|y)$ cannot be calculated analytically, its expression simplifies
in the small $\alpha$ limit. With the change of
variable $x=y(1-\alpha u)$, we obtain the following distribution of
$u$ at leading order in $\alpha$:
\beq
\rho_{\rm eq}(u)\approx \frac{1}{4u^3}e^{-\frac{1}{2u}},
\eeq
from which we deduce $\<u\>\approx 1/2$ and thus
\beq\label{accuracy_switch}
P_{\rm
  wrong}=\frac{1-\<xy\>}{2}= \frac{\alpha \<u\>}{2} \approx \frac{\alpha}{4}.
\eeq
This analytic prediction agrees very well with the simulation, see
Fig.~\ref{fig:accuracy}a. This error is estimated assuming that the
system has adapted, so long after the adaptation transient, $t\gg
t_A$.

However, while $P_{\rm wrong}$ is the error long after the adaptation
transient, it is not equal to the average error including during the
adaptation transients. To get that average error, we can approximate
the error to $1$ for $0<t \leq t_A$, and to $P_{\rm  wrong}$ for
$t>t_A$. Since the average time between switches is $2\tau$, we
obtain:
\beq
\begin{split}
  P_{\rm  wrong,av} &\approx \left(1-\frac{t_A}{2\tau} \right)P_{\rm  wrong} + \frac{t_A}{2\tau} \\
  &\approx \frac{\alpha}{8}
  \left[\ln^2(\alpha)+(2\gamma_{\rm e}-2\ln(2)+1)\ln(\alpha)+C+2\right].
  \end{split}
\eeq

Note that this scaling of the error with $\alpha$ is smaller than for
the OU process \eqref{accuracy_ou}. In the switching process, in the $\alpha\ll 1$ limit the system adapts almost perfectly
to the value of $y$ after a transient lag.
By contrast, in the OU process the state variable changes constantly,
causing a permanent lag in the adaptation, and thus a larger error.

{\ch
Fig.~\ref{fig:information}b shows how transmitted information depends
on $\alpha$, in particular how it quickly converges to the maximal
transmissible information as $\alpha$ goes to $0$.

  \begin{figure*}[t!]
\centering
\includegraphics[width=1.0\textwidth]{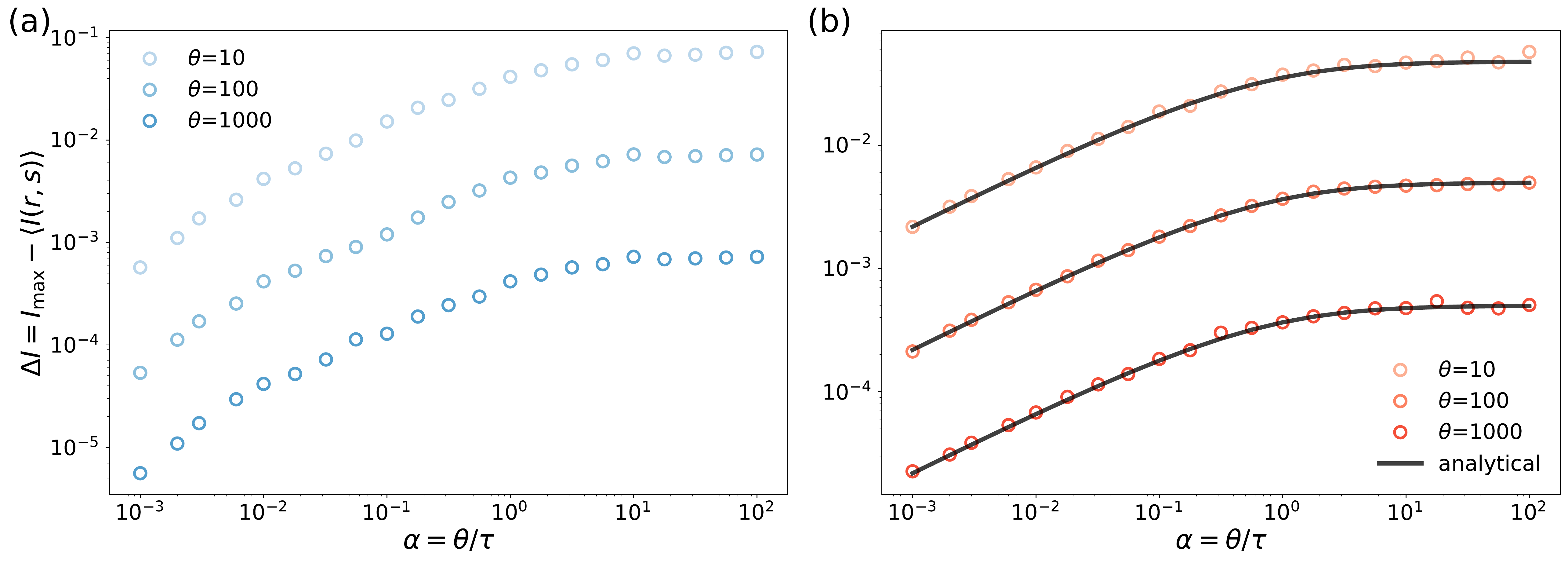}
\caption{\ch
\textbf{Average information transmission as a function of the
  adaptation parameter $\alpha=\theta/\tau$.} Difference $\Delta I$ between the
mutual information \eqref{eq:Is} averaged over time, $\<I(s,r)\>$, and
the channel capacity $ I_{\rm max}=\frac{1}{2}\ln[2\pi e(r_{\rm max}/\sigma_\epsilon)^2]$, for (a) the
telegraph process, and (b) the Ornstein-Uhlenbeck process.
Circles are from numerical simulations, and the line is the analytical prediction given by \eqref{eq:IOUavrg}.}
\label{fig:information}
\end{figure*}
}

\begin{figure*}[t!]
\centering
\includegraphics[width=1.0\textwidth]{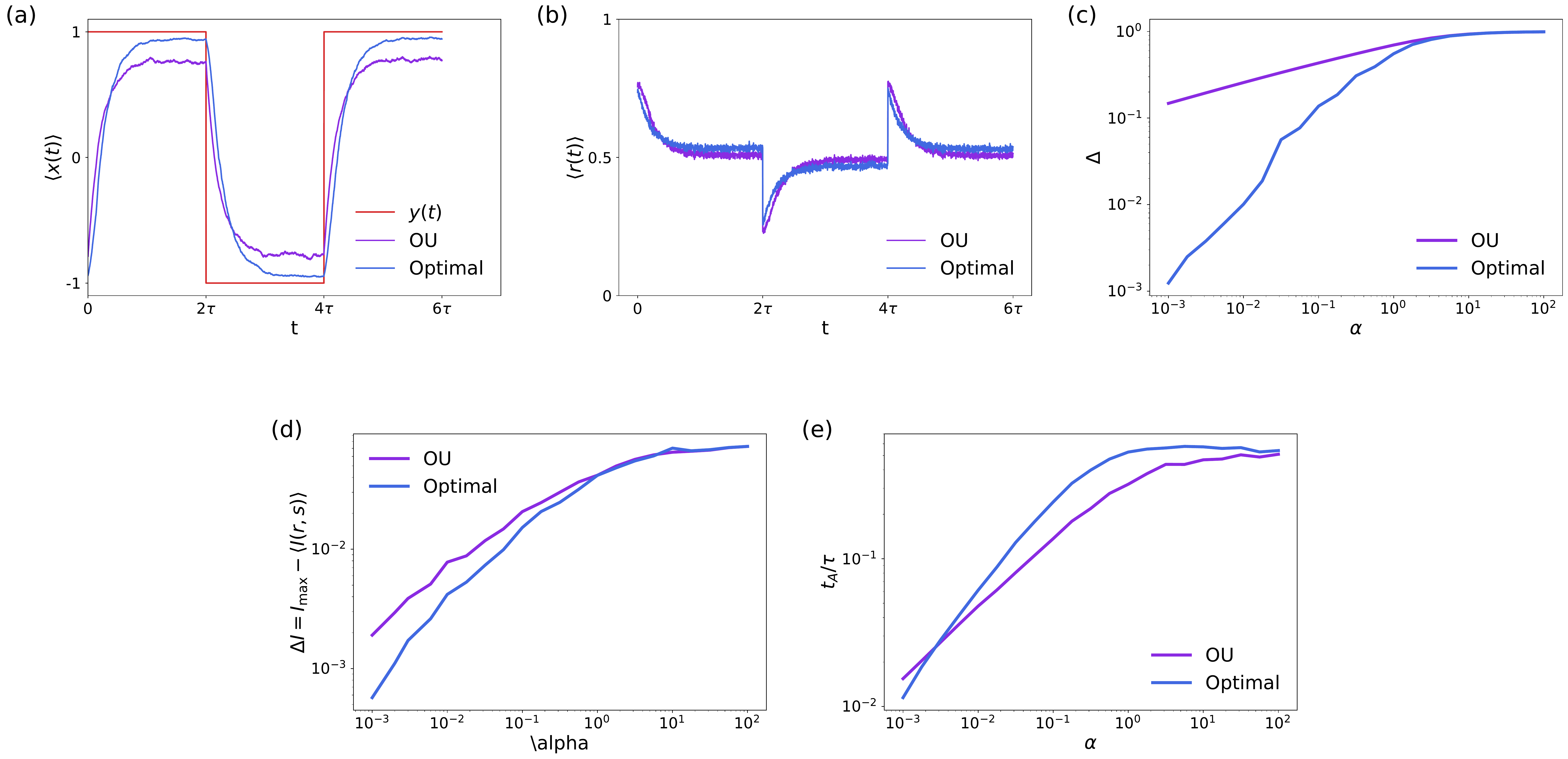}
\caption{
\textbf{Optimal adaptation vs OU adaptation in Telegraph process.}
\textbf{a)} Example of adaptation of a telegraph process given by $y$, red line, using an optimal estimator, blue line, and a (non-optimal) OU estimator, purple line.
\textbf{b)} Average response of the adaptive system based on a gain
function derived from an optimal estimator (blue), or from the OU estimator (purple).
\textbf{c)} Accuracy of adaptation measured by $\Delta = \sqrt{\langle
  (y-x)^2\rangle}$ using the optimal (blue) and OU (purple) estimator.
{\ch {\bf (d)} Information transmission using the optimal (blue) or OU
  (purple) estimators.
  }
\textbf{e)} Rescaled adaptation time as a function of the control
parameter $\alpha$ using optimal (blue) and OU (purple) estimators.
Note the crossover between the OU and the Optimal around
$\alpha=10^{-3}$ as the fastest adapting estimator.
}
\label{fig:opt_vs_ou}
\end{figure*}

\section{Non-optimal adaptation}\label{sec:maladapt}

So far we have assumed that the sensory system knows perfectly the statistics of the environmental dynamics, including its time scale $\tau$, mean and, variance, and type of dynamics.
In realistic situations however, the system may not know the
statistics of the environment precisely, or may have evolved to best
respond to stimuli with partiular statistics (e.g. natural ones),
while the experimental stimulus was designed using artificial statistics
(e.g. mean switching between two values). In this section we explore the impact of maladapted (non-optimal) adaptation dynamics, where the inference system makes wrong assumptions about the environmental dynamics.

\subsection{Misrepresented dynamics}
We first consider adaptation when the sensory systems misrepresents the
nature of the dynamics. We assume that the true dynamics follows a switching
mean (telegraph process), while the Bayesian system assumes an
Ornstein-Uhlenbeck dynamics. This leads to integrating the OU
dynamics \eqref{ou_eq}, but with $y$ switching between $-1$ and $+1$
with rate $(2\tau)^{-1}$.

Such an adaptation scheme is illustrated Fig.~\ref{fig:opt_vs_ou}a-b.
The error (in the sense of the OU process, $\Delta^2=\<(x-y)^2\>$) may be calculated
analytically and is identical to the case where $y$ would actually
follow a OU process with the same time constant, Eq.~\ref{accuracy_ou}. This is due to the
fact that in both cases the solution to \eqref{ou_eq}  can be
formally written as $y(t)=\int^t dt'\, e^{-(t-t')/\tau'}(\tau'/\theta)
u^2 s(t')$, with $\tau'=\tau/\sqrt{1+2/\alpha}$, and that
both the telegraph and the OU process have the same
first two moments \eqref{eq:moments}.

However, it does
much worst that the optimal adaptation scheme \eqref{continuous_x}, as
can be seen from Fig.~\ref{fig:opt_vs_ou}c, and confirmed by the
scaling at small $\alpha$ of the adaptation time ($\alpha\ln^2\alpha
\ll \alpha^{1/2}$, see ~\eqref{eq:OUsmallalpha}). {\ch The impact of
  this non-optimal adaptation can also be seen, to a lesser extent, on
  mutual information (Fig.~\ref{fig:opt_vs_ou}d).}

For the same reason cited above, the adaptation time scale 
$t_A$ is identical to that of the OU process \eqref{tau_ou_full}.
We observe a
broad range of parameters where adaptation with the OU assumption is
{\em faster} than the optimal adaptation scheme
(Fig.~\ref{fig:opt_vs_ou}e). We can understand this intuitively by
noting that the OU assumption assumes constant change, and thus will
be quicker to react to sudden changes. The flipside is that even when
the mean is stable between switches, it overestimates the uncertainty and
tends to conservatively adapt less well  (Fig.~\ref{fig:opt_vs_ou}a
and b).

\subsection{Wrong parameters}
We next consider the case where the type of dynamics is correctly
assumed by the Bayesian system, but its parameters are wrongly
estimated. The assumed values of the two parameters are off by a
factor $r$ and $l$ respectively: $\tau_{\rm assumed}=r\tau$,
$\theta_{\rm assumed}=l\theta$. The resulting adaptation dynamics for
both processes
\eqref{continuous_x} and \eqref{ou_eq} are modified by substituting these
assumed values:
\begin{align}
  \frac{dx}{dt} & = -\frac{x}{r\tau} + \frac{1-x^2}{l\theta} y + \frac{1-x^2}{l\sqrt{\theta}} \xi,\\
  \frac{d x}{dt} &= -\frac{\sqrt{1 + 2{r}/(l\alpha)}}{r\tau}x
                   +\frac{u^2}{l\theta} y +\frac{u^2}{l\sqrt{\theta}}
                   \xi,
\end{align}
for the telegraph and OU processes respectively, with $u^2=2/(1+\sqrt{1+2r/(l\alpha)})$.
On the other hand, the dynamics of $y$ (Eq.~\ref{eq:signal}, and switching with rate
$(2\tau)^{-1}$ for the telegraph process and Eq.~\ref{eq:ou} for the
OU process) are unchanged, with the true parameters $\tau$ and $\theta$.

For the OU process the problem is still solvable and the accuracy
can be calculated:
\beq
\Delta^2=u^2 \frac{1+lr+\beta (lr+r+l-1)+\beta^2(l-r)}{2l(1+\beta r)},
\eeq
with $\beta=(1+2r/(l\alpha))^{-1/2}$, which reduces to
\beq
\Delta^2\sim \frac{1+lr}{2\sqrt{lr}}(2\alpha)^{1/2}\geq (2\alpha)^{1/2}
\eeq
in the well-adapted regime $\alpha\ll 1$.
The adaptation time can similarly be computed:
\beq
\frac{t_A}{\tau}=\frac{\beta r}{1-\beta r}\ln \frac{2}{1+\beta r}.
\eeq
which scales as $t_A\sim (r\tau l\theta/2)^{1/2}$ in the $\alpha\ll 1$
regime, i.e. the geometric mean of the two assumed time scales, as in
the optimal case. In the $\alpha\gg 1$ regime, the
adaptation time scales like $\tau$ (also like the optimal case):
$t_A\sim \tau [r/(r-1)]\ln[(1+r)/2]$.
These results indicate that
while underestimating the noise and switching period lead to faster adaptation, doing so hurts the
accuracy of the inference.

For the telegraph process, the equilibrium accuracy may be computed using the same
method as described earlier:
\beq
P_{\rm wrong}=\frac{\alpha l}{4r}.
\eeq
For the adaptation time, we obtain
\begin{align}
\frac{t_A}{\tau} &\sim \frac{\alpha
                   l^2}{2(l-1)}\ln(1/\alpha)\quad\textrm{for }l>1\\
  & \sim \alpha^l \frac{l^2 \Gamma(l+1)}{2(1-l)}{\left(\frac{l^2}{2r}\right)}^{l-1}
   \quad\textrm{for }l<1\\
  & \approx \frac{\alpha}{4}\left[\ln^2(\alpha)+(1+2\gamma_{\rm
    e}-2\ln(2)+\ln(r))\ln(\alpha)+\mathcal{O}(1)\right]\nonumber\\
  &\qquad\textrm{for }l=1,
\end{align}
where $\Gamma(x)$ is the Gamma function.

Underestimating the switching rate ($r<1$) allows a slightly faster adapation,
but at the cost of a lower equilibrium accuracy. Likewise,
overestimating the noise ($l>1$) also speeds up the response, but again at the
expense of accuracy.

\section{Adaptation to varying variance}
\label{sec:variance}
\subsection{Optimal adaptation dynamics}

We now turn to the case of variance switching, where the sensory system tries to evaluate the variance of a random stimulus of fixed mean. The variance follows a telegraph process, alternating a between high-value $\sigma_h$ and a low-value $\sigma_l$.
There are many examples of this type of adaptation \cite{frazor2006local, rieke2001temporal, smirnakis1997adaptation, fernandez2010gain} and it has been much studied experimentally \cite{bialek2006efficient, brenner2000adaptive, fairhall2001multiple, lundstrom2008fractional, wark2009timescales, deweese1998asymmetric}.
As in the case of adaptation to the mean, we start with discrete time.  At each time step, the variance switches with probability $k$.
This defines a correlation time $\tau = 1/\ln(1-2k)$ of the underlying telegraph process.
The stimulus is $s_n=s_0+y_n\eta_n$, where $y_n$ is the varying variance (equal to $\sigma_h$ or $\sigma_l$), and $\eta_n$ is normally distributed ($\langle \eta_n \rangle = 0$ and $\langle \eta_n \eta_{n'} \rangle = \delta_{nn'}$). We set $s_0=0$, $\sigma_l=1$ and $\sigma_h=r$ without loss of generality. The posterior probability of being in the low-variance phase given previously observed stimuli, $P_n^l=P(y_n=\sigma_l|s_{j\neq n})$ is then given by the recursive relation obtained from \eqref{fullbayes}:
\begin{equation}
P^l_n = {\left[{1 + \frac{1}{r}\frac{k P^l_{n-1} + (1-k)(1 - P^l_{n-1})}{(1-k)P^l_{n-1} + k(1 - P^l_{n-1})} e^{-\frac{s_n^2 ({r^{-2}}-1)}{2}}}\right]}^{-1}.
\label{prob_discrete_var}
\end{equation}

A major difference with the case of adaptation to the mean is that this equation does not admit a well defined continuous-time limit. Intuitively, this is because learning the amplitude of white noise is much faster than learning its mean. During any observation time $\Delta t$, the observer has access to an infinite number of independent samples. However, each sample is infinitely noisy. For the mean, these two infinities compensate exactly, but for the variance the high value of the noise is not an issue since the goal is to estimate its magnitude. As a result, in the continuous-time limit, the system can adapt instantly to the variance, as it receives an infinite amount of signal. A solution to this problem could be to replace the white noise in the stimulus by a colored noise with a finite correlation time scale $\tau_s$, so that the number of independent samples during time $\Delta t$ scales as $\Delta t/\tau_s$. However, keeping with discrete time has similar effect with $\tau_s=\delta t$.

\begin{figure*}[htbp]
\centering
\includegraphics[width=1.0\textwidth]{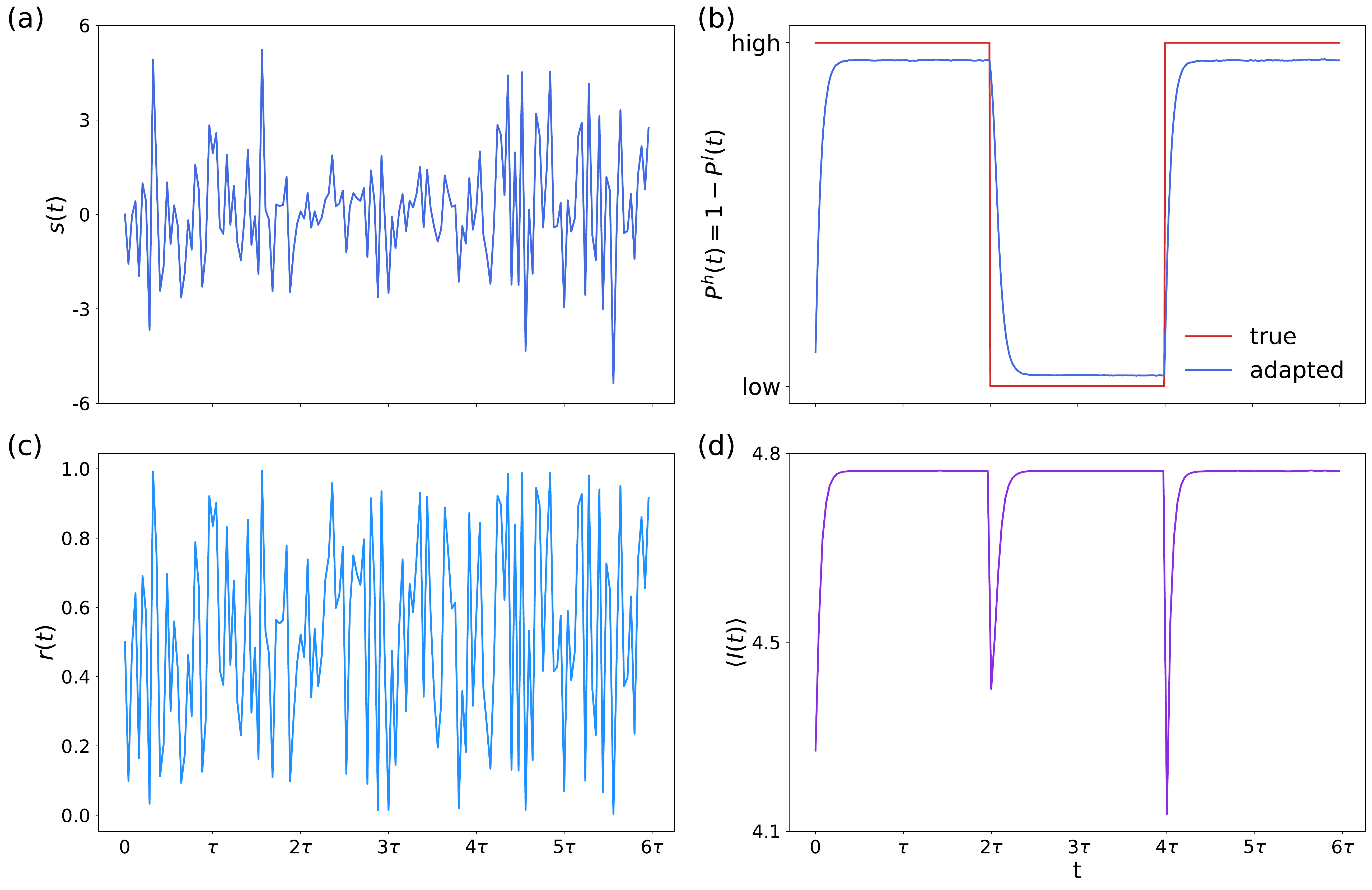}
\caption{
\textbf{Example of the variance adaptation process.}
\textbf{(a)} Typical stimulus received by the sensory system, made of normally distributed signals with swithcing variance.
\textbf{(b)} True value of the variance (red) versus assumed one (blue) based on past signals.
\textbf{(c)} Mean response obtained by the optimizing the expected information transmission based on the posterior distribution of variances.
Switching is hardly detectable in that response, as it affects transiently the variance but not the mean of the response.
\textbf{(d)} Average information transmitted per time bin, in bits ($\sigma_{\epsilon} =0.01$). Note the asymmetry between the two types of switches.
}
\label{fig:var_example}
\end{figure*}

\begin{figure*}[t!]
\centering
\includegraphics[width=1.0\textwidth]{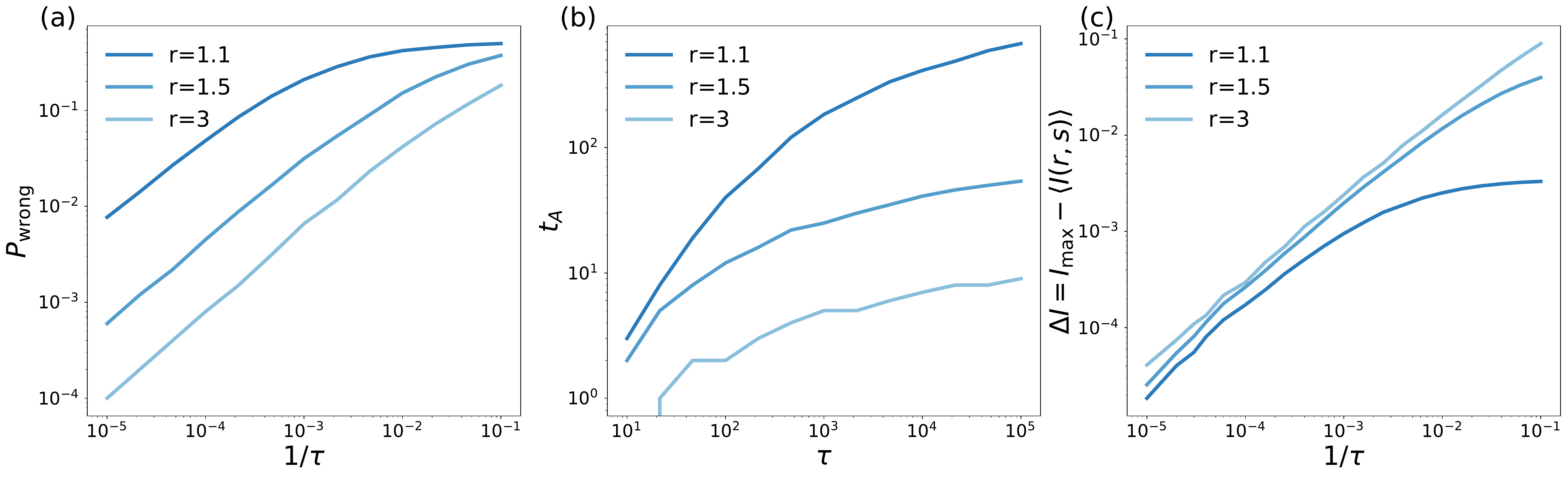}
\caption{
\textbf{Accuracy and speed of adaptation to variance switching.}
\textbf{(a)} Accuracy of the optimal adaptation dynamics for three different values of the variance ratio $r=\sigma_h/\sigma_l$, as a function of the inverse of the switching timescale $\tau$, which plays a similar role as $\alpha$ in mean switching.
The lower $r$ the larger the estimationg error.
\textbf{(b)} Adaptation time as a function of the switching timescale $\tau$. The dependence is linear for very short times, then displays strong sub-linear behaviour.
{\ch {\bf (c)} Information transmission as a function of $\tau$. Note that while a large ratio between the two variances induces a lower error in (a) (because they are easier to distinguish), this error has a larger impact on information transmission, so that information loss is overall larger.}
}
\label{fig:var_stat_fpt}
\end{figure*}

\subsection{Speed and accuracy of variance adaptation}
The dynamics of optimal adaptation can be studied by simulating \eqref{prob_discrete_var} numerically. Fig.~\ref{fig:var_example} shows an example of input signal, inferred variance, optimal output response and information transmission as a function of time.

The dynamics are characterized by an asymmetry in the adaptation to the two states (Fig. \ref{fig:var_example}b).
This is due to the fact that the high-variance state is capable to produce signals of low amplitude with high probability, while the low-variance state is very unlikely to produce large amplitudes. When in the low-variance state, the system receives few misleading signals, while in the high-variance state, frequent low-amplitude signals confound the posterior.
The flipside is that the system adapts faster to switches towards the high variance state, as already noted \cite{deweese1998asymmetric}, because that state produces unlikely signals under the low-variance hypothesis.
This asymmetric behaviour is reflected in the dynamics of information transmission, which experiences a stronger drop after a low-to-high than high-to-low transition (Fig. \ref{fig:var_example}d). Notably, these predictions agree with experiments in the fly visual system, where this asymmetry in the loss of information transmission following variance switches was observed \cite{Fairhall2001}.

As in the case of adaptation to the mean, we can define the accuracy of adaptation as the probability that the posterior is wrong, and the adaptation time $t_A$ as the delay $n$ maximizing $\<y_{n_0}P_{n_0+n}^l\>$. In Fig. \ref{fig:var_stat_fpt} we plot these two quantities, {\ch as well as the transmitted information, given by \eqref{eq:Is}}, as a function of the two parameters of the model: the switching rate $\tau$, and the variance ratio $r=\sigma_h/\sigma_l$. The error decreases with the switching period $\tau$, as the system has more time to adapt, as well as with the variance ratio $r$. The trend here depends also on the value of $r$.
The adaptation time varies sublinearly with $\tau$. It is approximately linear at small $\tau$, and levels off without saturating at larger $\tau$. The larger the variance ratio $r$, the narrower the linear region. {\ch The loss in information transmission due to imperfect adaptation, $\Delta I$, follows the same dependency on $\tau$ as the error. However, its dependency with respect to the variance ratio $r$ is reversed. The higher that ratio, the easier it is to distinguish between the two stimulus statistics, but errors are more costly in terms of information transmission because of the larger discrepancy between the assumed and actual dynamic range. Conversely, smaller differences between the two environments lead to less inefficiency in their encoding, despite a higher error in distinguishing them.}

Experiments in the fly \cite{Fairhall2001b} and vertebrate \cite{wark2009timescales} visual system suggest a linear relationship between the adaptation time and the switching timescale, although this linear assumption was not compared against alternative scalings. Our theory predicts that such a linear regime is only expected when the switching period is relatively short compared to the typical time between independent samples, which is also the regime where adaptation is poor.

\section{Entropy production as a signature of adaptation}
\subsection{Motivation and definition}
So far we have studied the adaptation dynamics of idealized sensory systems,
focusing on the trade off between precision and speed. However, these
notions require prior assumptions about what adaptation should 
look like and do. In the example of switching between environmental
states, the adaptation time course is expected to look like relaxation
dynamics. By contrast, for a continuous change such as OU, adaptation is much
harder to see by eye. Can we measure adaptation in a way
that makes minimal assumptions about what the response is adapting to,
or about the nature of the environmental changes?

To define such a quantity, we exploit the connection between
adaptation and the dissipative dynamics of a thermodynamic system
under an external drive \cite{crooks1999entropy,Lan2012}, as
suggested by the form of the adaptation dynamics
\eqref{continuous_x}-\eqref{ou_eq}, where the adaptation variable $x(t)$ is driven out of
equilibrium by an external stimlus $y(t)$.
In this type of dynamics, the importance of non-equilibrium effects is commonly measured by the
rate of entropy production \cite{lebowitz1999gallavotti}, which quantifies the irreversibillity of
the observed joint time courses of $x(t)$ and $y(t)$, and is defined
formally as the relative entropy between the forward and backward
trajectories of the dynamics.
Our goal is to calculate that entropy production in the systems
studied so far, and to relate this quantity to the degree of adaptation.

If $P(\Omega_0^N)$ is the probability of a given trajectory $\Omega =
\{z_0,..., z_N\}$  in discrete time, where $z=(x,y)$ is the variable
describing the state of the system, the entropy production of the
trajectory $\Omega$ is {\ch defined by \cite{Seifert2019}:
\begin{equation}\label{eq:EP}
    \mathcal{S}^N_{\mathrm{tot}}(\Omega) =  \ln{\frac{P(\Omega_0^N)}{P(\mathcal{R}(\Omega_0^N))}}
\end{equation}
where $\mathcal{R}(\Omega)$ is the time-reversal of $\Omega$.
While in early work entropy production was defined in
  terms of heat loss and dissipation, and shown to be equal to
  \eqref{eq:EP} in the case of Markovian dynamics with local detailed
  balance \cite{crooks1999entropy,seifert2005entropy}, here we take \eqref{eq:EP} as a definition, following the
  modern view of stochastic thermodynamics.}
For a Markov system, one has $P(\Omega_0^N)= p(z_0)\prod_{n=0}^{N-1}
w_{z_{n+1}z_n}$, where $w_{zz'}$ is the transition probability from
$z'$ to $z$, so that at steady state the average entropy production
per time step reads:
\beq\label{eq:deltaS}
\frac{1}{N}\mathcal{S}^N_{\mathrm{tot}}(\Omega)
\xrightarrow{N\to\infty}\delta S = \sum_{zz'} w_{zz'}p_{z'}\ln\frac{w_{zz'}}{w_{z'z}},
\eeq
where $p_z$ is the steady state distribution of $z$ defined by the
implicit equation: $p_z=\sum_{z'}
w_{zz'}p_{z'}$. Taking the continuous limit, we can define an entropy
production rate, $\dot S=\delta S/\delta t$, as $\delta t\to 0$.

When the statistics of stimuli do not vary and are reversible in time,
no adaptation occurs in an optimal estimator, and we expect the
temporal statistics of the response to be temporally reversible. When
the statistics of stimuli change (abruptly or continuously), the
response statistics are transiently poorly adapted and require time to
relax to its efficient encoding state, similar to the relaxation of
the equilibrium of a thermodynamic system after a change in an
external control parameter (temperature, displacement, force, etc.) thus leading to the production of entropy.

The measure of adaptation through entropy production provides a means
to detect and quantify adaptation in any sensory
system, in a parameter-free manner and without having to know the fine
details of the encoding strategy. In the following we quantify
the production of entropy in the models of mean and
variance adaptation studied above.

\begin{figure*}
\centering
\includegraphics[width=1.0\textwidth]{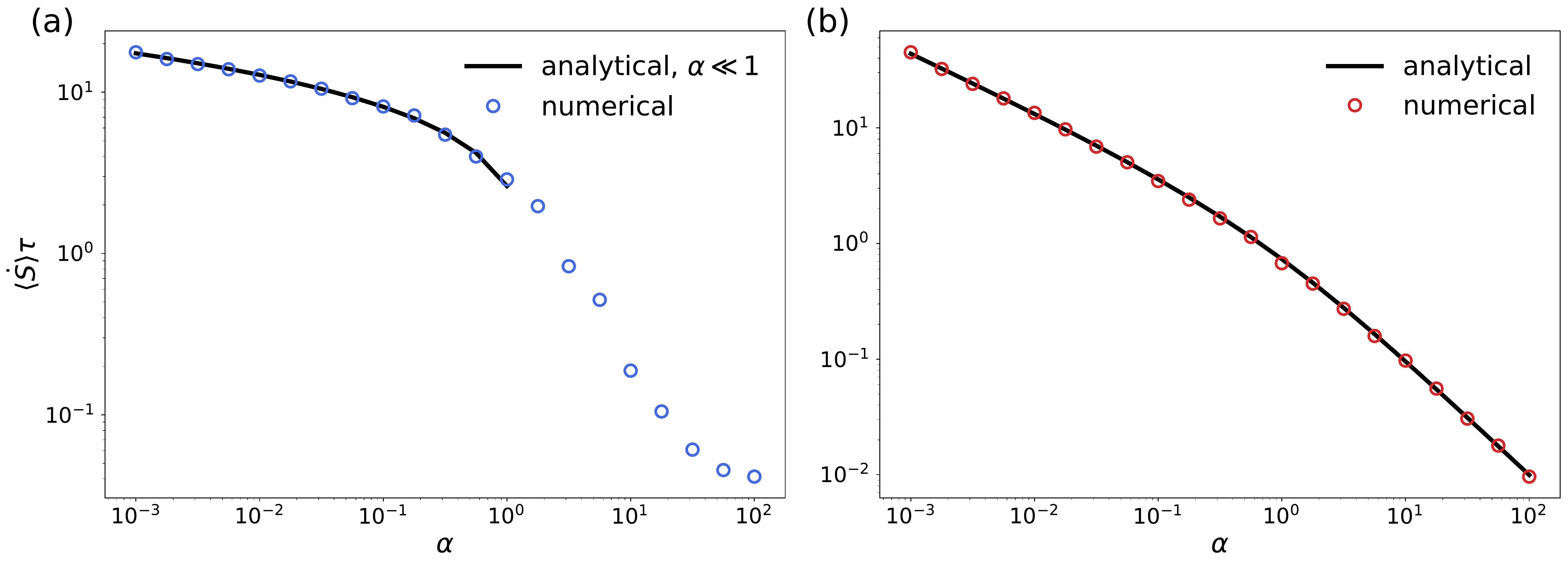}
\caption{
\textbf{Entropy production in mean adaptation process.}
Entropy production rate in the \textbf{(a)} telegraph and \textbf{(b)}
Ornstein-Uhlenbeck models of mean adaptation.
Points are obtained by the numerical simulations, while the curves are
analytical predictions given by \eqref{ent_mean_switch} for the
telegraph case and \eqref{ent_mean_ou} for the OU case. The lower the
$\alpha$ parameter, the better the adaptation, and the larger the
entropy production, which diverges with $\alpha\to 0$.
}
\label{fig:mean_ent}
\end{figure*}

\subsection{Mean switching}
We start with the OU process, for which entropy production may be
calculated analytically. We start with the Gaussian transition
probabilities, with $z'=(x(t),y(t))$ and $z=(x(t+\delta t),y(t+\delta t))$:
\begin{align}
  w_{zz'}&=P(y_{t+\delta t}| y_t) P(x_{t+\delta t}|x_t, y_t )\\
P(y_{t+\delta t}| y_t) & \propto \exp-\frac{\left[y_{t +\delta t} - y_t\left(1 - \frac{\delta t}{\tau}\right)\right]^2}{4\frac{\delta t}{\tau}}, \\
    P(x_{t+\delta t}|x_t, y_t ) &\propto \exp-\frac{\left[x_{t +\delta t} - x_t\left(1 - \frac{\delta t}{\tau'}\right) - \frac{\delta t}{\theta}y_t\right]^2}{ 4 D_x \delta t},
\end{align}
with $D_x=u^4/2\theta$. Plugging these expression into \eqref{eq:deltaS} gives:
\begin{equation}
 \delta S = \frac{\sqrt{1+2/\alpha} -1}{\tau}\frac{C(\delta t) - C(-\delta t)}{2 D_x}
\end{equation}
where we recall $C(t)=\<y(t_0)x(t_0+t)\>$.
Expanding at small $\delta t$ yields the entropy production rate:
\begin{equation}
  \dot{S}\tau = \sqrt{1+2/\alpha} - 1
\label{ent_mean_ou}
\end{equation}
which diverges as $\propto \alpha^{-1/2}$ in the regime of good
adaptation ($\alpha\ll 1$), and decays to zero when signal is too poor
to allow for adaptation ($\alpha\gg 1$).

To calculate the entropy production during the period of a switch in
the well-adapted regime ($\alpha\ll 1$), we
may exploit the analogy with statistical thermodynamics. After each
switch, the energy landscape changes from $U_+(x)=-\ln \rho_{\rm
  eq}(x|y=1)$ to $U_-(x) =-\ln \rho_{\rm eq}(x|y=-1)$ (and
vice-versa), which are given by the steady-state distributions
\eqref{eq:eq}.
During the adaptation transient, the
system relaxes to its new equibrium, lowering its energy by
dissipation. The heat thus dissipated is exactly equal to the total
entropy produced during this transient. Assuming that the system has
completely adapted to the previous epoch prior to each switch (which
is valid for $\alpha\ll 1$), we may write the entropy production rate
as the mean entropy production per switch, divided by the average time
between switches:
\beq
\dot S  = \frac{1}{2\tau}\left[\<U_-(x)\>_{x\sim\rho_{\rm eq}(\cdot,y=+1)}-\<U_-(x)\>_{x\sim\rho_{\rm eq}(\cdot,y=-1)}\right],
\eeq
\beq
\dot S\tau =\left\<\ln\frac{1+x}{1-x}\right\>_{x\sim\rho_{\rm eq}(\cdot,y=+1)}.
\eeq
In the small $\alpha$ regime, we can compute this expression at leading order:
\begin{equation}
\dot{S}\tau  = 1-\gamma_{\rm e} +2\log(2) -\log(\alpha),
\label{ent_mean_switch}
\end{equation}
which again diverges for $\alpha\to 0$, but this time logaritmically. In the
$\alpha\to \infty$ limit, the system never adapts, and the dynamics
are in effective equilibrium, $\dot{S}\tau \rightarrow 0$.

\subsection{Variance switching}
\begin{figure}
\centering
\includegraphics[width=\linewidth]{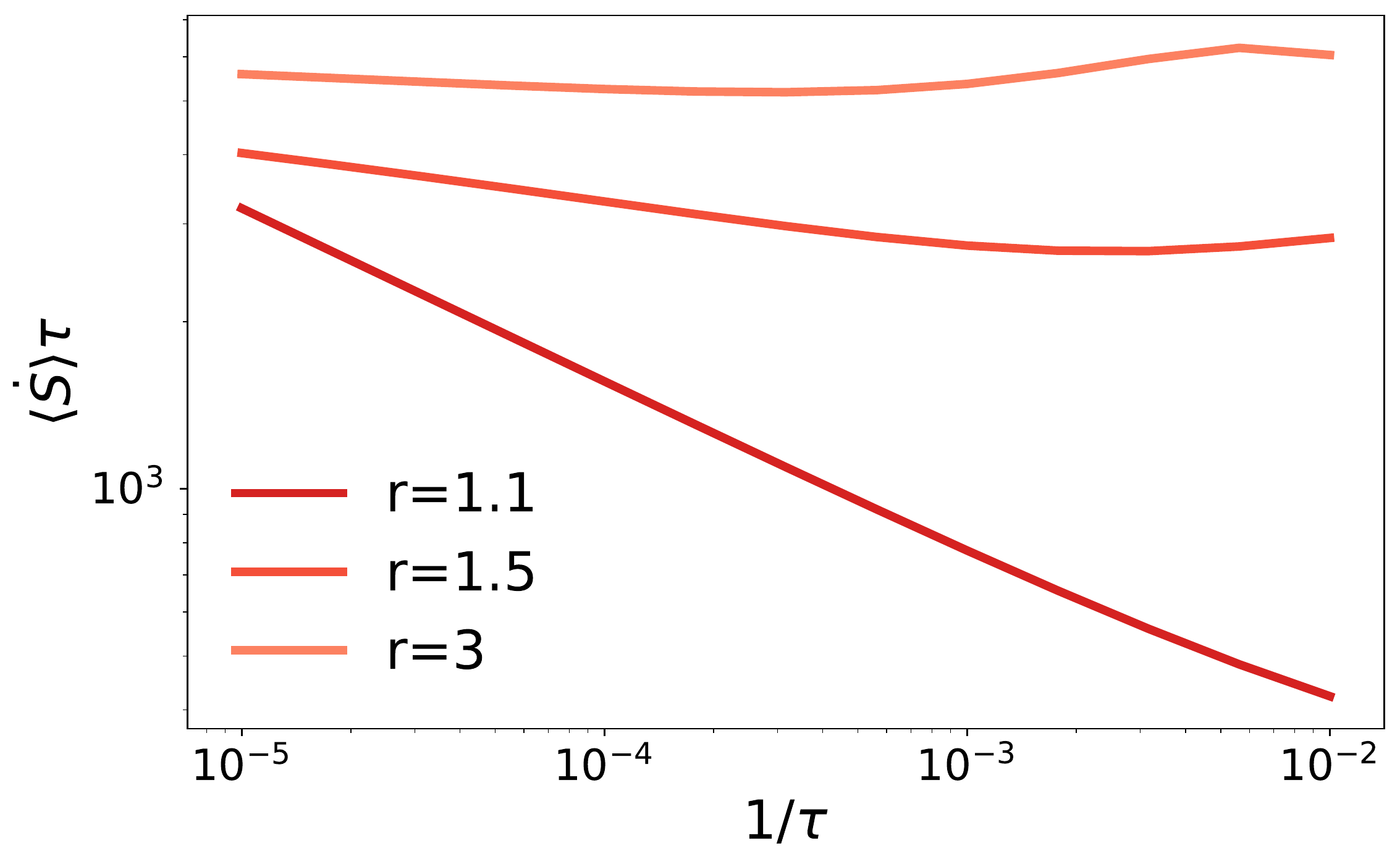}
\caption{
\textbf{Entropy production in adaptation to variance switching.}
Average entropy production rate $\delta S \tau$ as a function of the inverse
correlation time for three different values of the variance ratio $r=\sigma_h/\sigma_l$.
}
\label{fig:tau}
\end{figure}

Since we have no analytical solution or continuous-time limit for the
case of a switching variance, we recourse to numerical estimates of \eqref{eq:deltaS}.
The results, shown in Fig.~\ref{fig:tau}, paint a similar picture as
for the mean switching case. When the switching period is large, the
system is well adapted, spending most of its time in the 
equilibrium state, but dissipating heat during each adaptation
transient. The average amount of entropy production per switching
cycle seems to diverge for large $\tau$, and also grows with the
variance ratio $r$, as expected.

Taken together, these results suggest that entropy production provides
a good signature of adaptation. It is larger when adaptation is more
precise, and when the environemental changes causing adaptation are
more important.

\section{Discussion}

We have studied a theoretical scheme in which sensory adaptation is cast as an inference problem. As noted before \cite{deweese1998asymmetric,wark2009timescales,mlynarski2018adaptive}, the solution to this problem can be described in terms of Bayesian filters. In this work, we have focused on simple, solvable models of adaptation to sudden or continuous changes to the stimulus statistics. The resulting Bayesian adaptation dynamics couple the sensory response function to the state of the environment in a noisy manner, allowing for their study in terms of stochastic thermodynamics. This explicit analogy to non-equilibrium systems allowed us to relate adaptation and dissipation (entropy production) in a precice mathematical sense.

In the case of a switching stimulus mean, these dynamics reduce in the continuous time limit to coupled stochastic differential equation, which we can solve analytically. We computed the speed and accuracy of adaptation, and showed that they both depend on a single parameter $\alpha$ controling the ratio between the noise and the environmental time scale. When $\alpha$ is small, the system has enough time to adapt relative to the environment's speed of change. In this regime adaptation is accurate, meaning that the response function is optimally tuned to the current environment statistics. For large $\alpha$ however, the environment changes too quickly for the system to garner information, and the response does not adapt, instead ignoring received signals and using an effectively constant response function ($x\approx 0$).

We find that, in the fidelity regime, the timescale of adaptation always scales {\em sublinearly} with the environmental timescale. However, the precise scaling depends of the precise model of the environment: logarithmic for abrupt changes in the stimulus mean (telegraph process), and square root for continuous ones (Ornstein-Uhlenbeck process). Since these are two extremes of one-dimensional stochastic process, we expect other types of environmental dynamics to have scalings that fall within that range.
Most experiments studying visual adaptation have reported linear scalings in adaptation to both the mean and variance \cite{Fairhall2001b,wark2009timescales}, although alternative scaling laws were not tested. In our model, a strict linear scaling is approached only in the limit of low fidelity, where adaptation is very poor. This suggests that these sensory systems are actually not optimal for these simplified stimulus statistics, but have instead evolved under other constraints, such as metabolic or biophysical constraints, and for efficiency under a broad range of complex stimulus contexts.

Our results on mean switching mostly hold true in the case of variance switching, including the sub-linear scaling of adaptation time with environmental timescale.
In that case, the continuous time limit is not straightforward. We thus focused on the dynamics in discrete time using numerical simulations.
We found an asymmetry in the adaptation dynamics, with a larger drop in information transmission following an increase of variance than a decrease, consistent with experimental observations in the fly visual system \cite{Fairhall2001}.

Optimality is only a guiding principle, and not an assumption for what the system does. We also studied the effect of optimizing response properties when stimulus statistics are misrepresented. We found that while assuming continuous change is more conservative, the corresponding adaptation strategy fares just as well when the actual dynamics are discontinuous. Mischaracterizing the parameters of the environmental statistics affects the precision and speed of adaptation, tipping the balance between the two, but not their scaling with $\alpha$.

To explore the impact of adaptation on the coding strategy, we assumed that the system changes its response function to optimize the expected information transfer.
This relies on the implicit assumption that the sensory system also outputs the adaptation variable $x$ in addition to the response $r$. However, realistic coding strategies may combine the two together into a single response, as for instance in the case of variance coding, where the response encodes both stimulus fluctuations and its varaince through the mean reponse \cite{Fairhall2001}. How to efficiently compress these different signals into a single noisy response remains to be studied.

{\ch A main contribution of this paper is a mathematically precise link
between adaptation and dissipation.
This analogy} emerges from the form of the dynamical Bayesian equations \eqref{continuous_x}-\eqref{ou_eq}, where adaptation resembles relaxation to equilibrium.
We showed that a high degree of adaptation, as determined by its speed and accuracy, can be quantified by the degree of deviation from equilibrium and reversibility.
This measure of adaptation through entropy production could give a systematic way to detect and quantify adaptation in any sensory system, in a parameter-free manner and without having to know the fine details of the encoding strategy. {\ch In particular, it could be applied to recordings of {\em populations} of neurons in a variety of sensory systems in response to changes in the stimulus statistics, using increasingly available multi-electrode recordings \cite{Marre2012a} or 2-photon calcium imaging techniques \cite{Stringer2019,Privat2019}. Adaptation has mostly been studied at the level of single neurons, because of its evident manifestation in terms of spiking rates of individual neurons. However, more subtle adaptive changes could occur in the way multiple neurons encode information collectively through their interactions or shared variability, which would be missed by traditional methods. For instance, Hopfield networks \cite{Hopfield1982a} encode information in the collective state of many neurons, and it has been argued that similar principles may govern the encoding of visual information in the retina \cite{Schneidman2006}. The signatures of entropy production proposed here could be useful for detecting adaptation in such combinatorial codes.
}

Other sources of irreversibility, including in the stimulus statistics themselves, or through the inherent irreversibility of the biophysical processes that implement the response, may confound this analysis and would have to be corrected for.
Still, the relation between adaptation, time and accuracy provides a testable hypothesis that could be explored in future experiments. 
It should also be emphasized that this dissipation may not directly correspond to actual energy consumption in the system, although it always provides a lower bound. In neural systems, energy consumption from electrical activity far outweighs dissipation estimated by the irreversibility of measurable quantities. On the other hand, the two may be surprisingly close for molecular systems. Previous work by Lan et al. \cite{Lan2012} has used entropy production to estimate energy consumption in the context of adaptation in {\em E. coli} chemotaxis, showing that the system realizes a near-optimal trade off between energy, speed, and accuracy. By contrast, our proposal is to use entropy production as an intrinsic definition of adaptation in sensory systems, independently of energetic considerations.

Our framework could be expanded to account for ``meta-adaptation'' \cite{robinson2016meta}, which describes the possibility that the system dynamically learns the hyperparameters of the stimulus statistics ({\em e.g.} $\theta$, $\tau$, and $\sigma$) from the past stimulus. {\ch A similar form of meta-adaptation has been proposed and explored in the context of evolutionary adaptation \cite{Katz2016}, where microbial populations are assumed not only to adapt their composition to environmental changes, but also to the statistics with which that environment changes, using a similar strategy of Bayesian filtering as employed here.}
Combining all hyperparameters into a collective variable $\Theta$, we could write a similar recursive equation to \eqref{fullbayes} assuming a Markovian dynamics for $\Theta$:
\begin{equation}\begin{split}
   P(y_n, \Theta_n| s_{j \le n}) =& \frac{1}{\Omega}P(s_n|y_n) \bigg[\sum_{y_{n\shortminus1}, \Theta_{n\shortminus1}} P(y_n|y_{n\shortminus1},\Theta_{n-1})\\
     &\times P(k_n|\Theta_{n\shortminus1})P(y_{n\shortminus1}, \Theta_{n\shortminus1}|s_{j <n})\bigg].
    \label{meta}
\end{split}\end{equation}
In this scheme the sensory system learns both the parameters of the stimulus statistics as well as the hyperparameters that govern their own dynamics. Such meta-adaptation is needed to explain how sensory systems dynamically adapt their adaptation speed to the environmental timescale \cite{fairhall2001multiple,wark2009timescales}, on a longer timescale than adaptation itself.
This hypothesis is consistent with the fact that changes in the world occur on many time scales \cite{Dong1995}, and is also suggested by the wide range of timescales in visual adaptation even at the level of the retina, from seconds to hours. Future experimental and theoretical work should determine the relevance of this theory in biological sensory systems.

\bigskip

{\bf Acknowledgements.} The authors were supported by grants ANR-17-ERC2-0025-01 ``IRREVERSIBLE'' and ANR-19-CE45-0018 ``RESP-REP'' from the Agence Nationale de la Recherche.

\bibliographystyle{pnas}

\appendix


\section{Mean first passage time after a switch}
\label{app:fpt}
We start with \eqref{fpt} operate the change of variable
$x=1-\alpha/(2w)$, $x'=1-\alpha/(2v)$. In the limit $\alpha\ll 1$, we
assume that $w$ and $v$ are small for the dominating contribution of
the integral. Because of the $(1-x^2)$ prefactor
in the attraction term towards $y$, the dynamics of $x$ is initially
``stuck'' in the previous belief, $x\sim 1$. Once it reaches a value
$1-\mathcal{O}(1)$, it quickly (in time $\sim \theta$) reaches $0$,
before continuing on to its next state $\sim -1$.

The change of variable gives:
\beq\label{eq:fpt2}
\begin{split}
\mathcal{T}(x_0)\sim & \frac{\theta}{2}\int_{\frac{\alpha}{2}}^{w_0}dw\,
{e^{\frac{w^2}{w-\frac{\alpha}{4}}}}\frac{w-\frac{\alpha}{4}}{w^2} \int_w^{+\infty}
  dv\,e^{-\frac{v^2}{v-\frac{\alpha}{4}}}\frac{v^2}{(v-\frac{\alpha}{4})^3},
  \end{split}
\eeq
with $w_0=(\alpha/2)/(1-x_0)$ is of order $\mathcal{O}(1)$ in
$\alpha$. The internal integral in \eqref{eq:fpt2} can be rewritten as
the sum of three parts:
\beq\label{eq:i}
\begin{split}
&\approx \int_w^{+\infty} \frac{dv\,e^{-v}}{v}+\int_w^{+\infty} R(v)
e^{-v}{v} \\
&+ \int_w^{+\infty}
\frac{v^2}{(v-\frac{\alpha}{4})^3}e^{-v}\left(e^{-\frac{\alpha}{4}\frac{1}{v-\alpha/4}}-1\right)
\end{split},
\eeq
with $R(v)={v^2}/{(v-\frac{\alpha}{4})^3}-1/v$.

The first term in \eqref{eq:i} is by definition equal to the special function $E_1(w)\approx w-\gamma_{\rm
  e}-\ln(w)+\mathcal{O}(w)$ (related to exponential integral). Its contribution to \eqref{eq:fpt2} is
dominated by the behaviour at small $w\to \alpha/2$ where the integral
diverges when $\alpha\to 0$. As we will see, it also dominates the second and
third terms, yielding:
\beq
\begin{split}
\mathcal{T}(x_0) & \approx
  \frac{\theta}{2}\int_{\frac{\alpha}{2}}^{w_0} dw\,
    \ln(1/w)-\gamma_{\rm e}\frac{w-\frac{\alpha}{4}}{w^2}\\
  & \approx
  \frac{\theta}{4}\left[\ln^2(\alpha)+(2\gamma_{\rm
      e}-2\ln(2)+1) \ln(\alpha)+\mathcal{O}(1)\right].
  \end{split}
  \eeq

What remains to show is that the second and third terms in \eqref{eq:i} contribute at
most $\mathcal{O}(1)$ to the result.
The second term's contribution to \eqref{eq:fpt2} is bounded by:
\beq
\begin{split}
  &\frac{\theta}{2}e^{2w_0}\int_{\alpha/2}^{w_0}\frac{dw(w-\alpha/4)}{w^2}\left[\frac{8\frac{2w}{\alpha}-3}{2\left(2\frac{2w}{\alpha}-1\right)^2}+\ln\frac{\frac{2w}{\alpha}}{\frac{2w}{\alpha}-\frac{1}{2}}\right]\\
  &={\theta}{}\mathcal{O}(1).
  \end{split}
\eeq

As for the third term, it is approximately:
\beq
\approx -\frac{\alpha}{4}\int_{w}^{+\infty}
\frac{w^3}{\left(w-\frac{\alpha}{4}\right)^4}e^{-w}\propto \alpha\ln\left(\frac{1}{w}\right),
  \eeq
  and will thus contribute
  $\mathcal{O}(\theta\alpha\ln(1/\alpha))=\mathcal{O}(\theta)$ to the result.

\end{document}